\documentclass[12pt]{iopart}

\expandafter\let\csname equation*\endcsname\relax
\expandafter\let\csname endequation*\endcsname\relax

\usepackage{amsmath}
\usepackage{graphicx}
\usepackage[T1]{fontenc}
\begin{document}

\title[Modelling of spatial structure of divertor footprints caused by
RMP-mitigated ELMs]{Modelling of spatial structure of divertor footprints caused by edge-localized modes mitigated by magnetic perturbations}

\author{P Cahyna$^1$, M Becoulet$^2$, G T A Huijsmans$^3$, F Orain$^2$, J Morales$^2$, A Kirk$^4$, 
   A J Thornton$^4$, S Pamela$^4$, R Panek$^1$ and M Hoelzl$^5$}
\address{$^1$ Institute of Plasma Physics AS CR, Prague, Czech Republic}
\address{$^2$ CEA/IRFM, Cadarache, 13108 Saint Paul Lez Durance, France}
\address{$^3$ ITER Organization, Route de Vinon-sur-Verdon, CS 90 046,
  13067 St. Paul Lez Durance Cedex, France}
\address{$^4$ CCFE, Culham Science Centre, Abingdon, Oxon, OX14 3DB, UK}
\address{$^5$ Max Planck Institute for Plasma Physics, Boltzmannstr. 2, 85748 Garching, Germany}

\begin{abstract}
{
Resonant magnetic perturbations (RMPs) can mitigate the edge-localized modes (ELMs), i.e. cause a change of the ELM character towards smaller energy loss and higher frequency. During mitigation a change of the spatial structure of ELM loads on divertor was observed on DIII-D and MAST: the power is deposited predominantly in the footprint structures formed by the magnetic perturbation. In the present contribution we develop a theory explaining this effect, based on the idea that part of the ELM loss is caused by parallel transport in the homoclinic tangle formed by the magnetic perturbation of the ELM. The modified tangle resulting from the combination of the ELM perturbation and the applied RMP has the expected property of bringing open field lines in the same areas as the tangle from the RMP alone. We show that this explanation is consistent with features of the mitigated ELMs on MAST.
We in addition validated our theory by an analysis of simulations of mitigated ELMs using the code JOREK. We produced detailed laminar plots of field lines on the divertor in the JOREK runs with an ELM, an applied RMP, and an ELM mitigated by the presence of the RMP. The results for an ELM clearly show a high-n rotating footprint structure appearing during the nonlinear stage of the ELM, which is not present in the early stage of the ELM. The results for a $n=2$ RMP from the ELM control coils show the expected $n=2$ footprint structure. The results for the mitigated ELM show a similar structure, modulated by a higher $n$ perturbation of the ELM, consistent with our theory.
}
\end{abstract}
\maketitle

\section{Introduction}
Resonant magnetic perturbations (RMPs) are one of the promising methods of
edge-localized mode (ELM) control in future tokamak reactors~\cite{Evans2013S11}, which is
necessary as already for ITER
it is predicted that at full plasma current uncontrolled ELMs
can not be tolerated~\cite{0029-5515-54-3-033007}.  The RMPs either suppress the ELMs completely or
cause a change of the ELM character towards smaller energy loss and
higher frequency~\cite{fenstermacheriaea2010}. The latter effect is called ELM mitigation and may
be a viable option for ELM control on ITER if the reduction in peak heat
load at the divertor is sufficient to protect the plasma-facing
components. One major unknown parameter in the predictions of
tolerable ELM sizes is the ELM wetted area, which together with the
total ELM loss determines the peak heat load on the plasma-facing
components~\cite{0029-5515-54-3-033007,0029-5515-51-12-123001,Eich2011S856}. The compatibility of
mitigated ELMs with plasma-facing components is therefore not
only determined by the ELM energy loss reduction achieved, but also by
any changes of the wetted area accompanying the mitigation~\cite{0029-5515-51-12-123001,Eich2011S856}. As an
example, it was observed on MAST that mitigation achieves a less significant
reduction of peak ELM heat load than of the total ELM loss due to a
concurrent reduction of the wetted
area~\cite{0741-3335-55-1-015006,0741-3335-55-4-045007}. Similar
results were reported on JET with the carbon
wall~\cite{Jachmich2011S894}. On the contrary on JET with the
ITER-like wall the reduction of the peak heat load during ELM
mitigation was partly due to an increase of the wetted
area~\cite{0029-5515-53-7-073036}. The study
of spatial structure of ELM loads and its changes due to mitigation is
thus important for extrapolating ELM mitigation scenarios and still an
open area of research. 

One significant change of the spatial structure of ELM
loads on divertor during mitigation was observed on
DIII-D~\cite{0029-5515-49-9-095013}, MAST~\cite{Thornton2013S199} and NSTX~\cite{0741-3335-56-1-015005}: the power
is deposited predominantly in the footprint structures~\cite{Evans05} formed by the
magnetic perturbation. On MAST it is observed only at the early stage
of the ELM, though.  This effect may be one of the contributing
factors to the wetted area changes. Indeed on JET footprints of heat
flux are also seen during the mitigated ELM crash and were proposed as
the mechanism of the wetted area increase~\cite{0029-5515-53-7-073036}.
It also can provide a valuable insight
in the physics of ELMs and their mitigation. The effect suggests that
the phase of the ELMs becomes locked to the phase of the applied
RMP. The RMP has usually a low toroidal mode number though ($n=2,3$,
up to $n=4$ and $6$ on MAST) while the ELM may have a higher toroidal
mode number. On MAST the toroidal mode number of the ELM filaments was
found to be between 10 and 20 and, remarkably, was not affected by the
mitigation by $n=4$ or $n=6$
RMP~\cite{0741-3335-55-1-015006}. Although a high-$n$ ELM can
nonlinearly couple to a low $n$ perturbation~\cite{:/content/aip/journal/pop/20/8/10.1063/1.4817953}, it is not a priori clear
how this can lead to the ELM  producing a structure on the divertor which corresponds to a perturbation with a much lower $n$.

The goal of this paper is to present a simple mechanism which may in
some cases explain the observed peaking of ELM loads. It is assumed
that at least a part of the ELM energy loss is caused by parallel
transport along open, chaotic field lines in the homoclinic tangle
formed by the magnetic component of the
ELM~\cite{Huijsmans2013S57}. The simplest explanation would be thus
based on the analysis of the structure of the homoclinic tangle,
without considering MHD interactions of the ELMs with the applied
RMP. We will show that the structure of the homoclinic tangle itself
can indeed provide such an explanation. It is then needed to demonstrate
that the explanation remains valid when the MHD interaction of the ELM
with the RMP is taken into account. For this we will use the results
of simulations of RMP-mitigated ELMs with the nonlinear MHD code
JOREK. Finally we will discuss the relation of the mitigated ELM
features predicted by our model to the observations on MAST.

\section{Structure of the homoclinic tangle formed by the ELM and RMP}
\subsection{General description of the homoclinic tangle}
The homoclinic tangle is formed by two intersecting surfaces --- the
stable and unstable invariant manifolds of the
X-point~\cite{Evans05}. The manifolds themselves are formed by field
lines which approach the X-point asymptotically when followed in the
direction parallel or antiparallel to the field vector. In the
axisymmetric case they coincide and form the separatrix. Under the
influence of the magnetic perturbation they split and their distance
is in the first order approximation given by the Melnikov
integral~\cite{wiggins}. The usefulness of the concept of invariant
manifolds lies in the fact that they form the boundary of the region
of stochastic field lines which connect the divertor plates with the
plasma core. The intersection of the stochastic region with the
divertor plates manifests itself as a footprint receiving high
parallel heat fluxes and having long connection length. The shape of
the footprints can be calculated by tracing the connection length of
field lines starting at the divertor plates and representing the
result in a so-called laminar plot~\cite{WingenPoP09}. Alternatively the
Melnikov integral can be used to approximate the shape of the
manifolds. The Melnikov integral is a function of the field line label
at or near the separatrix, which we call the homoclinic coordinate
$h$~\cite{wiggins} and define it to be equal to the toroidal angle at
the outboard midplane and constant on field lines. As noted
in~\cite{EichPRL2003}, a radially displaced bundle of field lines at
the outboard midplane will show up as a spiralling structure on the
divertor, along a line of constant $h$. The displacement of the field
line is caused by the magnetic perturbation and the boundary of the
bundle is the invariant manifold. The Melnikov
integral $M(h)$ quantifies the displacement in flux coordinate of the invariant
manifold with respect to the original separatrix at the divertor
plates. The shape of the footprint boundary in the $h - \psi$ space is
therefore given by the equation $\psi = \max(M(h) , 0)$. $M(h)<0$
means that the manifold is inside the 
separatrix and therefore does not intersect the divertor. To know the
footprint shape on the divertor we perform the transform from the $h -
\psi$ space to the divertor coordinates $\phi - s$, where $s$ is the
radial distance along the divertor surface and $\phi$ the toroidal
angle. The manifold will then take an elongated shape along a spiralling curve of
constant $h$. More details can be found in~\cite{cahyna-sfp2013}.

The Melnikov integral can be calculated by integrating the radial
component of the perturbation along field lines on the unperturbed
separatrix. For our application it is crucial that the Melnikov
integral for a perturbation with a single dominant toroidal mode $n$
will also have a simple harmonic form: $M(h) = M_n \sin(nh +
h_0)$ which can be characterized only by the amplitude $M_n$, phase
$h_0$ and a toroidal mode number $n$. There is thus very little
freedom in the form of divertor footprints in a given equilibrium
magnetic field. The Melnikov integral is linear: for a linear combination
of perturbation fields, the resulting Melnikov integral is the
corresponding linear combination of their Melnikov integrals.

\subsection{Form of footprints of the combined ELM and RMP field}
\begin{figure}[htb]%
  \centering%
  \tiny{\includegraphics[width=52mm]{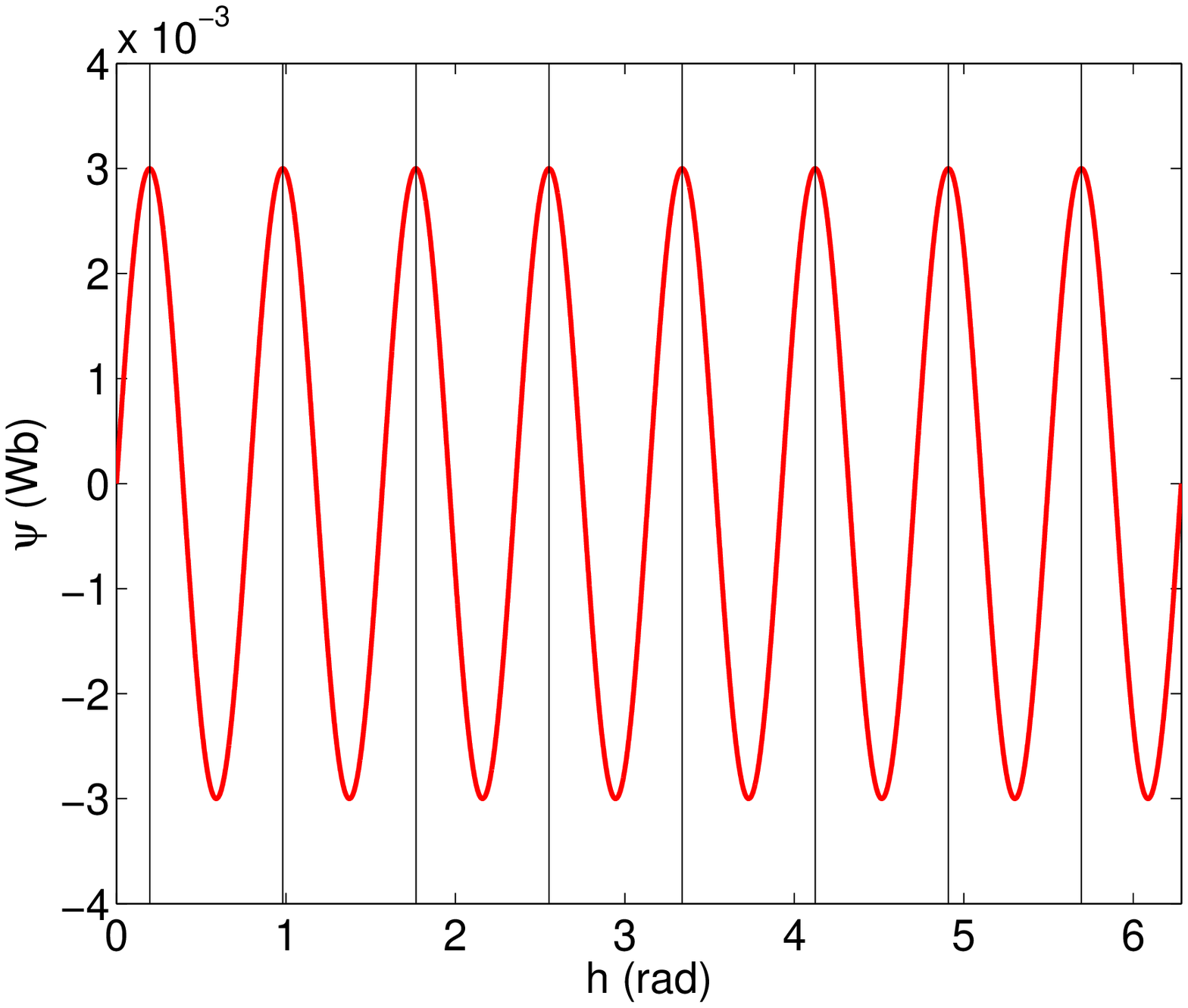}%
    \includegraphics[width=52mm]{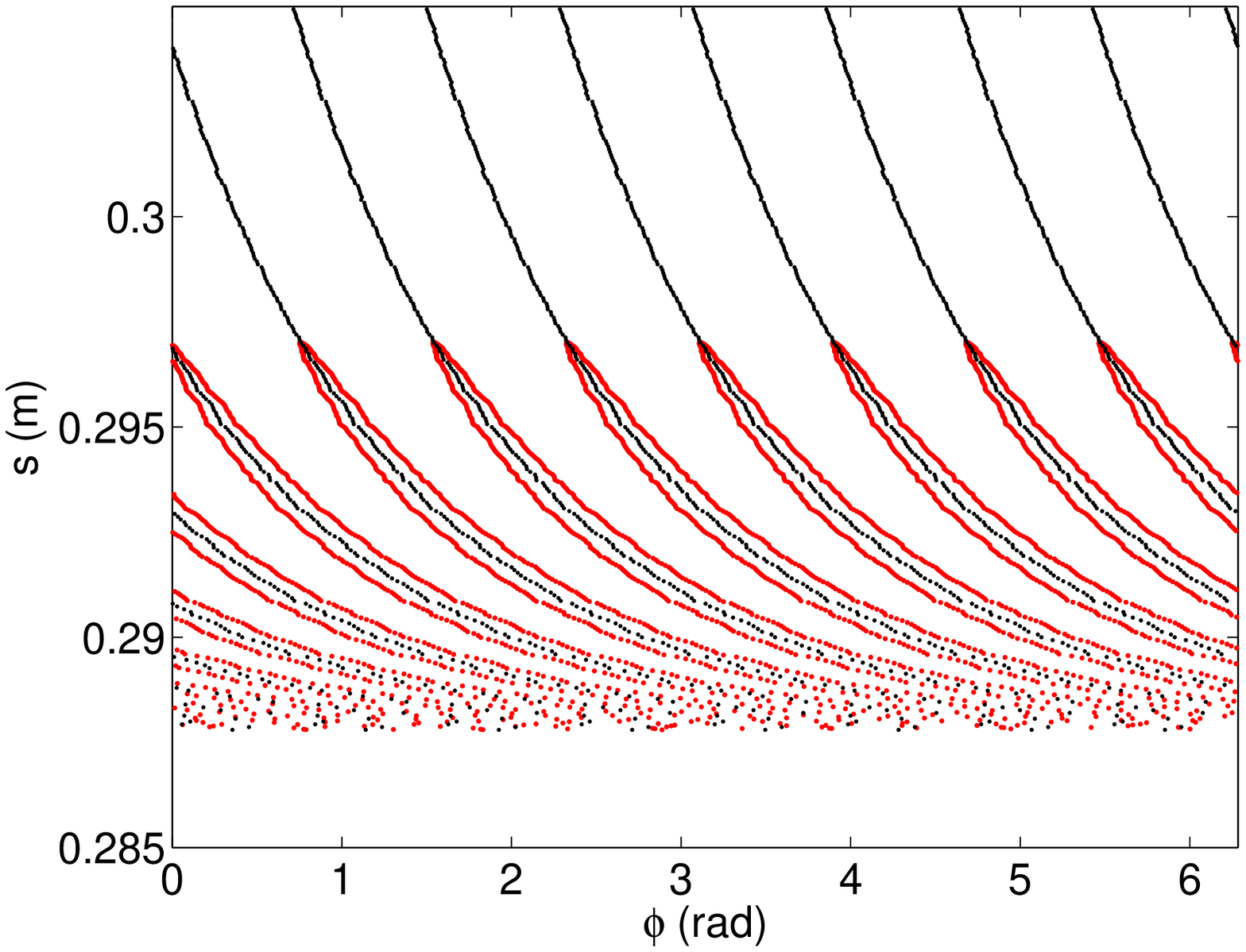}}%
  \caption{Footprints of a $n=8$ ELM in the in the $h - \psi$
    coordinates (left) and divertor coordinates (right). Black curves:
    coordinate curves of constant $h$.
  }%
 \label{fig:footprintn8}%
\end{figure}

We will use the Melnikov integral formalism to calculate the
footprints in the case of the ELM, in the case of the applied RMP and
in the case where the ELM is present together with the RMP field,
representing an mitigated ELM. We will represent here the ELM as a
magnetic perturbation with a higher toroidal mode number than the
RMP. For concreteness we will use $n=8$ for the ELM and $n=2$ for the
RMP. In the $h - \psi$ space the Melnikov integrals are simple
sinusoids: $M_{\text{ELM}}(h) = M_{\text{ELM}} \sin(8h +
h_{0\text{ELM}})$, $M_{\text{RMP}}(h) = M_{\text{RMP}} \sin(2h +
h_{0\text{RMP}})$.  The footprints of the ELM on the divertor (with
coordinates $s$ -- the radial distance along the plate and $\phi$ --
the toroidal angle) form a similar
pattern as the footprints of the RMP, only with a finer structure of
spirals. Figure~\ref{fig:footprintn8} shows the invariant manifold (red) with the curve of
constant $h$ (black, compare with fiures in~\cite{EichPRL2003}). Indeed, field line tracing in
a simple model of the ELM magnetic perturbation indicates~\cite{EichPRL2003} and
observations on ASDEX~Upgrade confirm~\cite{0741-3335-47-6-007} that the ELM create a spiral-shaped
heat load footprint which lies on a curve of a constant toroidal
angle when mapped to the midplane (called $h$ here), in agreement with our picture.

\begin{figure}[htb]%
  \centering%
  \tiny{\includegraphics[width=52mm]{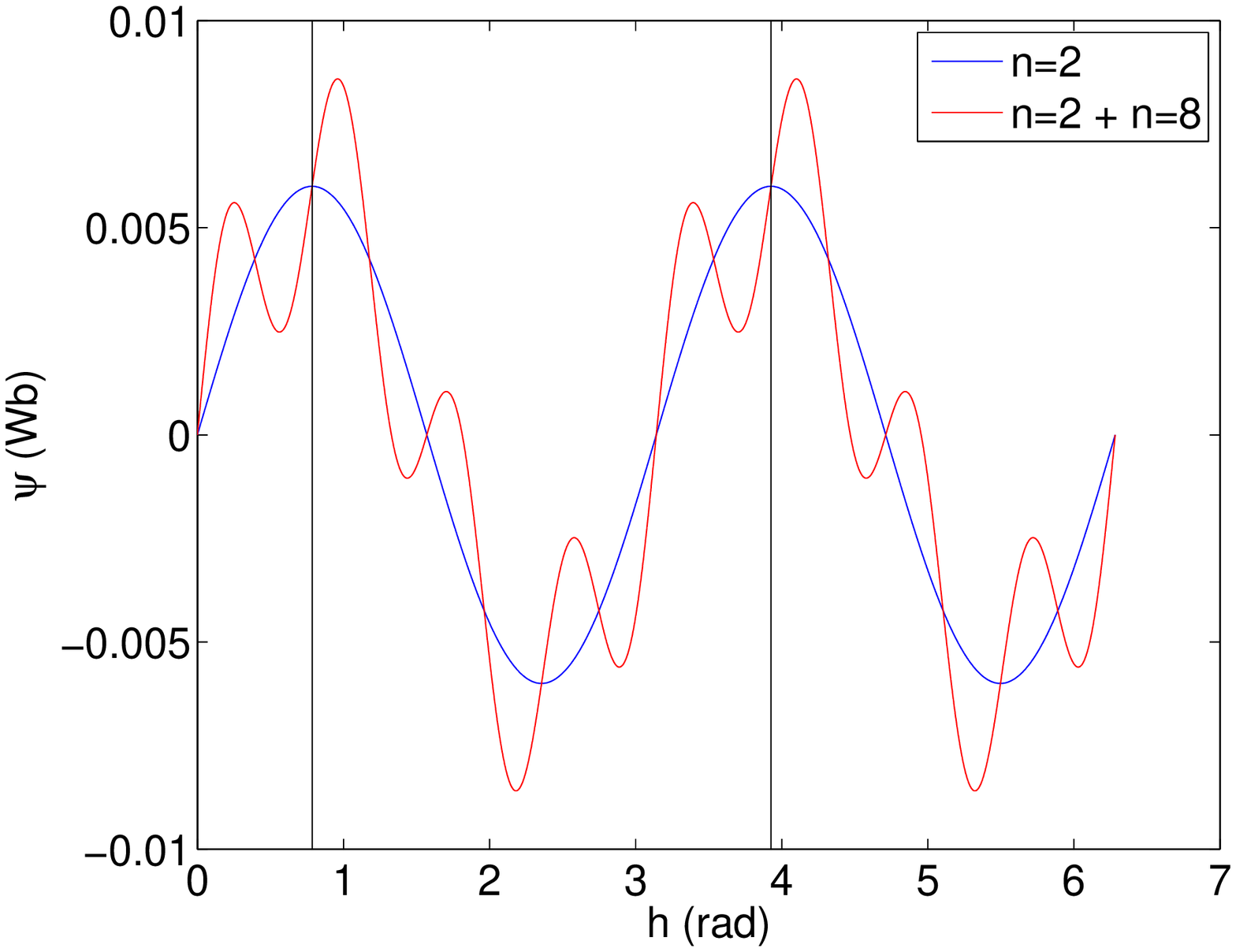}%
  \includegraphics[width=52mm]{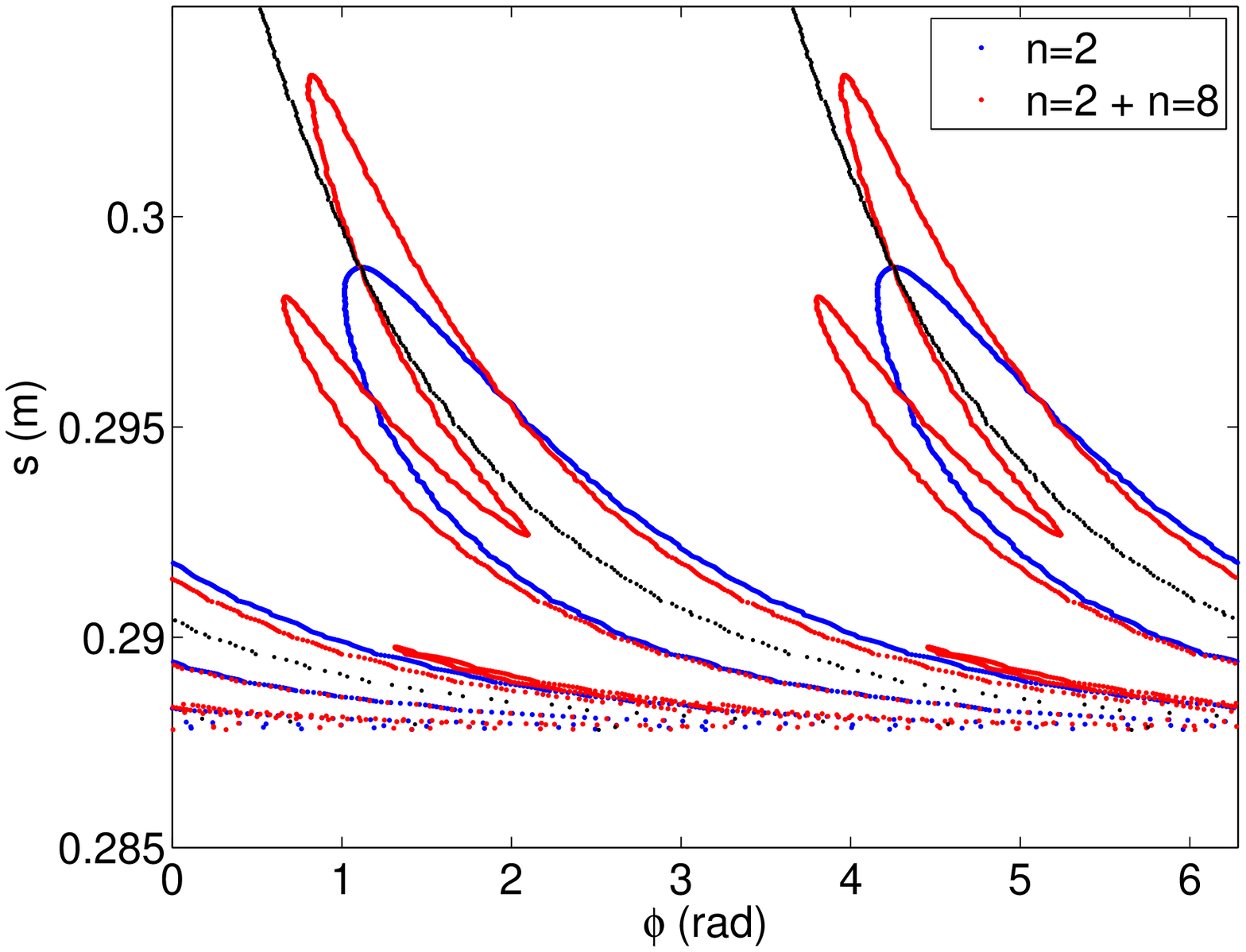}}%
  \caption{Footprints of a $n=2$ RMP (blue) and superposed with a $n=8$ ELM
    (red) in the $h - \psi$ coordinates
    (left) and the divertor coordinates (right).}%
 \label{fig:footprintn2n8}%
\end{figure}

For the combined RMP and ELM field we start with the simplest
assumption that the RMP does not have an influence on the ELM and the
total field is merely a linear superposition of the RMP and a natural
ELM: $M(h) = M_{\text{ELM}}(h) + M_{\text{RMP}}(h)$ (figure~\ref{fig:footprintn2n8}, left). The ELM will then
act as a carrier wave modulated by the RMP and its fine footprints
will be concentrated in the areas of the footprints of the RMP (figure~\ref{fig:footprintn2n8}, right). If we
average over multiple ELMs or over time while one ELM is rotating
(this can be represented by varying the phase $h_{0\text{ELM}}$), the
average footprint pattern of heat deposition will be similar to the
one formed by the RMP alone, in agreement with experimental
observations. This effect relies on the ELM perturbation amplitude to
be comparable or smaller than the RMP amplitude: $M_{\text{ELM}} \leq
M_{\text{RMP}}$, otherwise the RMP structure would be obscured by the
  larger ELM structure.

In practice we may expect the RMP to interact with the ELM, in
particular to decrease its amplitude (mitigation). This does not
significantly affect our reasoning and may  help to satisfy the
relation $M_{\text{ELM}} \leq M_{\text{RMP}}$ even in cases where it is
not satisfied for unmitigated ELMs.

\section{Verification of the model using MHD simulations}
\subsection{Form of footprints in the MHD simulations}
In the previous sections we developed an understanding of
the form of mitigated ELM divertor footprints based on a very simple
representation of the perturbed magnetic field. We shall now verify
the model in a more realistic magnetic field and a full 3-D
geometry. Such calculations were already performed
in~\cite{0029-5515-54-6-064011} where the ELM field was modelled using
a source formed of a set of equidistant current filaments. Here we
will use an even more realistic model: MHD simulations
of the ELM crash with the code JOREK~\cite{0741-3335-51-12-124012,:/content/aip/journal/pop/20/10/10.1063/1.4824820}, including RMPs. The simulations
were run for a JET-like equilibrium. The JOREK
simulations use a Fourier representation in the toroidal direction,
harmonics $n=0,2,4,6,8$ were used in the present simulations. The
applied RMP is here produced by the
error field correction coils (EFCCs) of JET, but only the dominant $n=2$
toroidal harmonic of their field is present (as the boundary condition for the
 $n=2$ toroidal harmonic of the simulated field), the higher $n$ sidebands that occur in
realistic coil designs are omitted for simplicity. More details on the simulations can be
found in~\cite{marinaiaea2014,marinaPRL2014}. The evolutions of
magnetic energies of the toroidal modes are shown in figure~\ref{fig:modes},
together with the time instants where the following figures are
taken.
\begin{figure}[htb]%
  \centering%
  \tiny{\includegraphics[height=45mm]{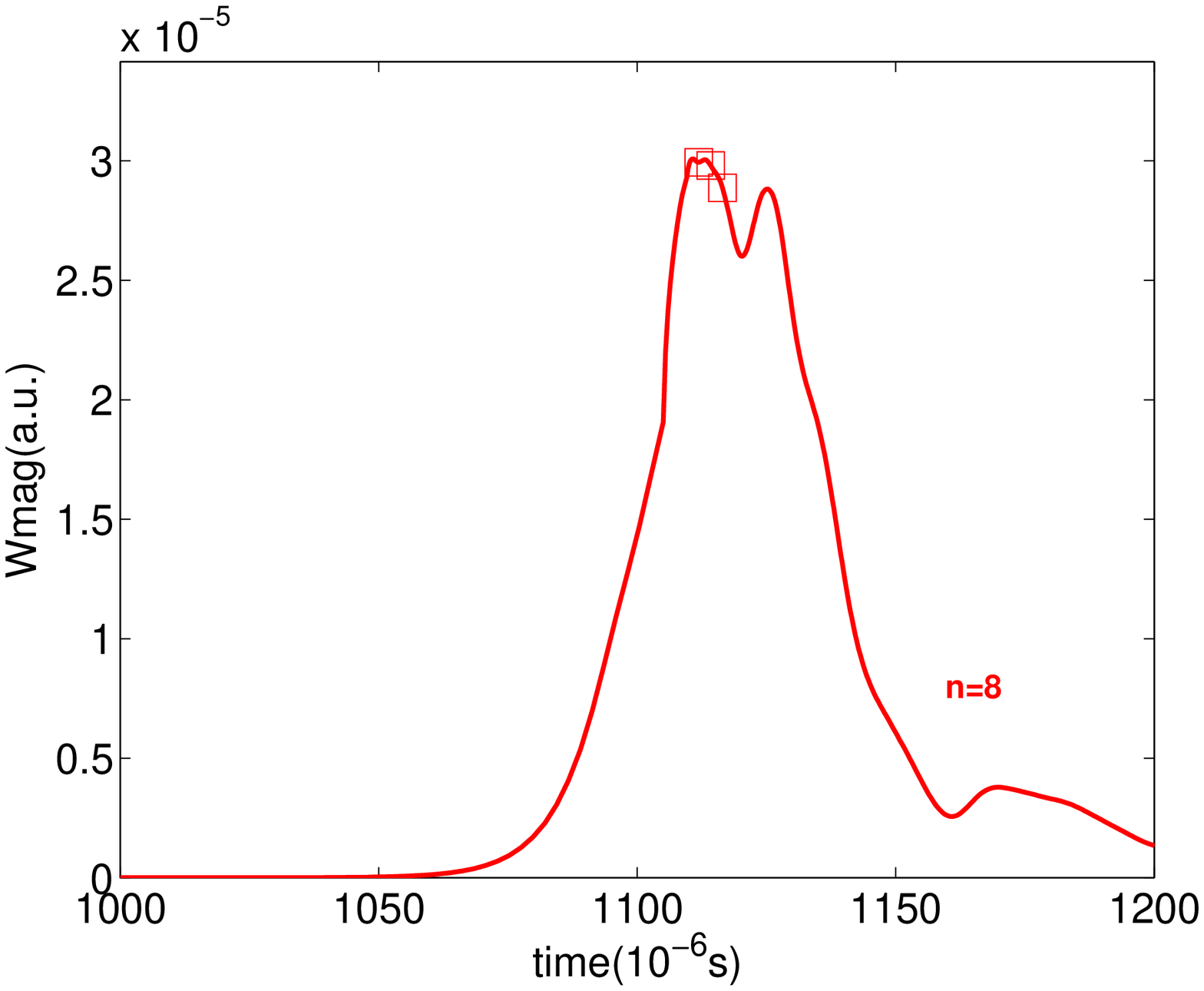}%
  \includegraphics[height=45mm]{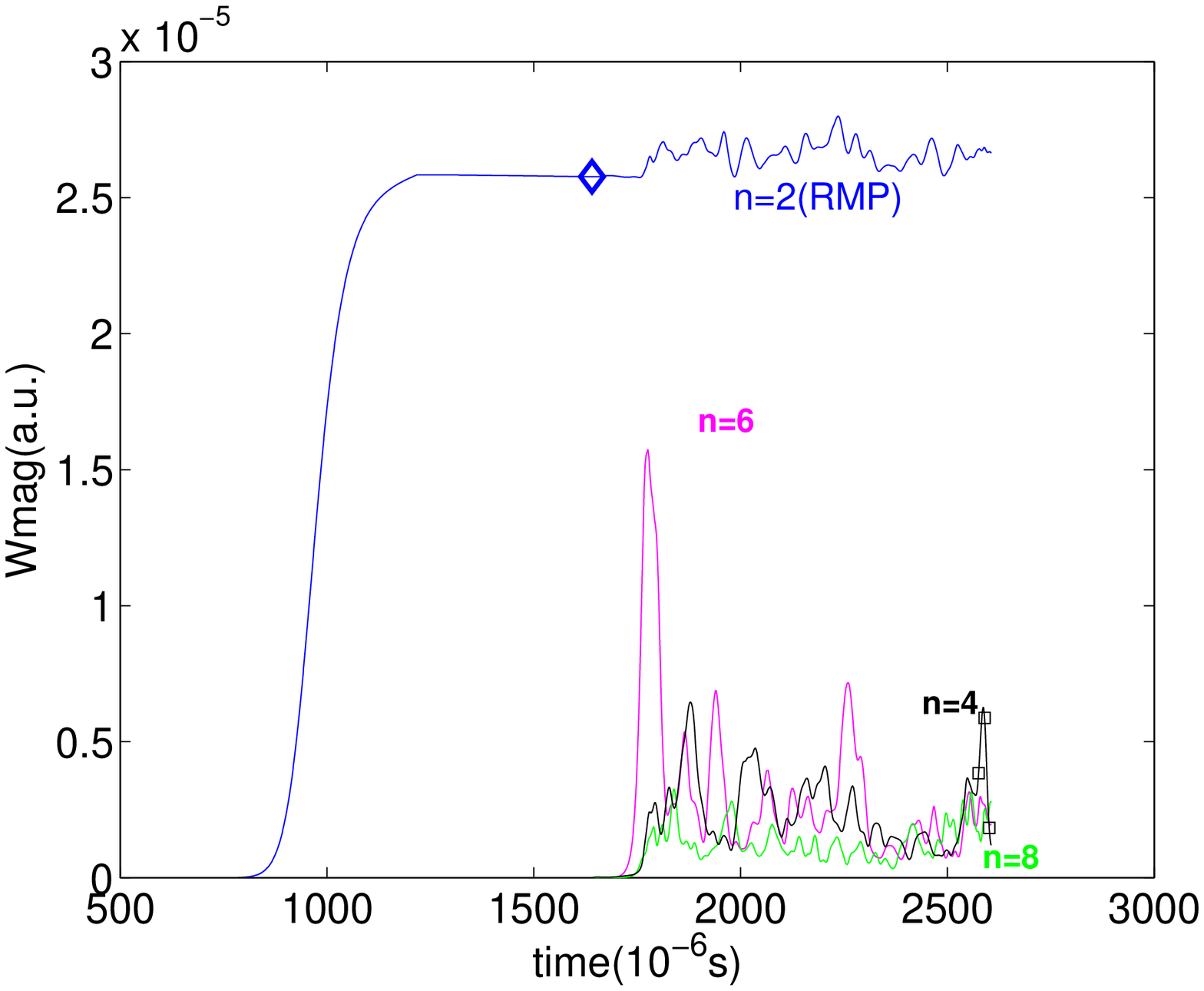}%
  \includegraphics[height=45mm]{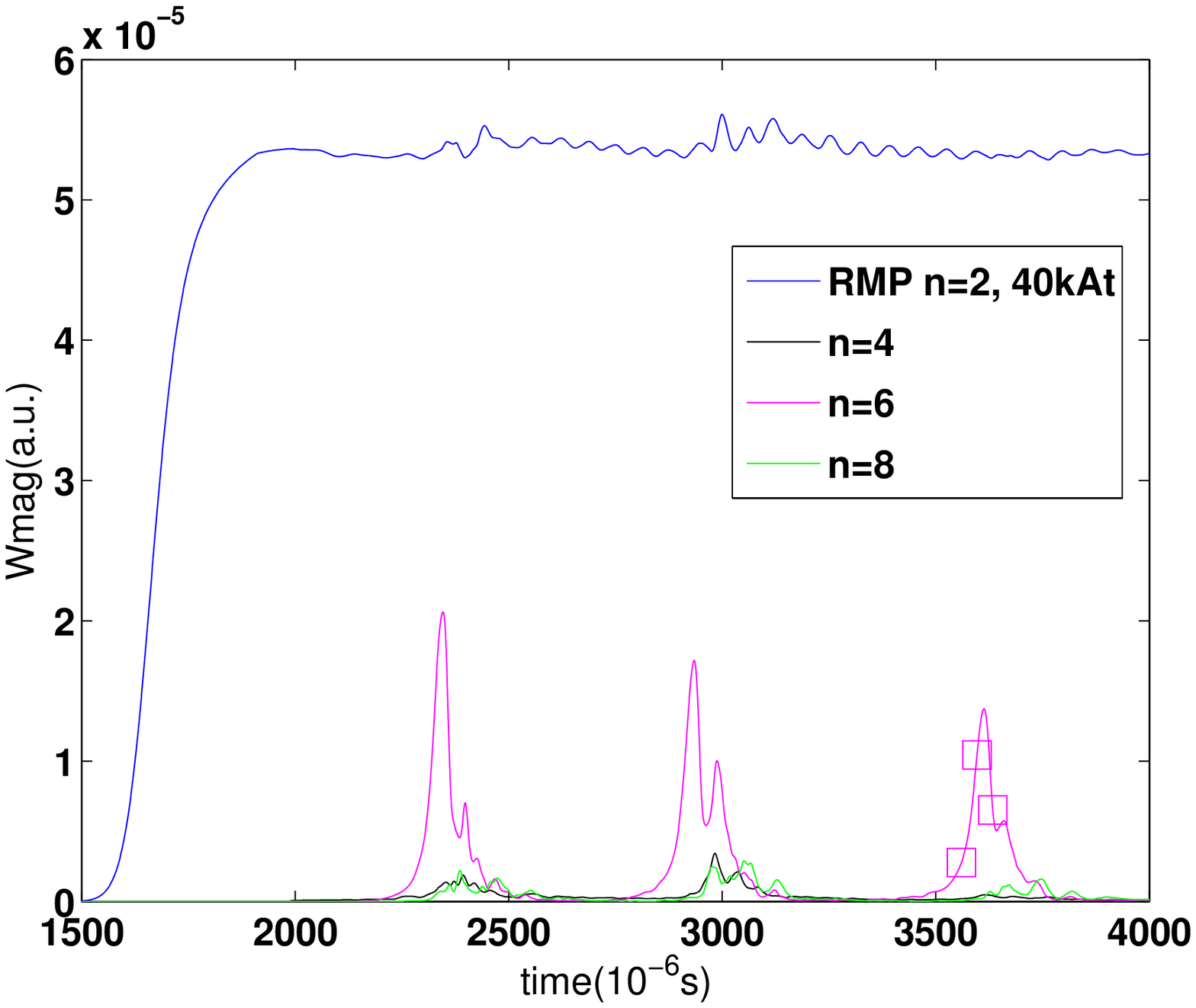}%
  }%
  \caption{Time evolution of the magnetic energies of the modes in the
  JOREK simulations used: ELM crash without RMP (left, corresponding
  to figure~\ref{fig:n8ELMonly}), $n=2$ RMP and
  mitigated ELMs (center, corresponding
  to figure~\ref{fig:ELMplusRMP}) and ELM cycles not mitigated by RMP
  (right, corresponding
  to figure~\ref{fig:ELMplusRMPn6}). The symbols show the time instants
  corresponding to the following laminar
plots.}%
 \label{fig:modes}%
\end{figure}
\begin{figure}[htb]%
  \centering%
  \tiny{\includegraphics[width=50mm]{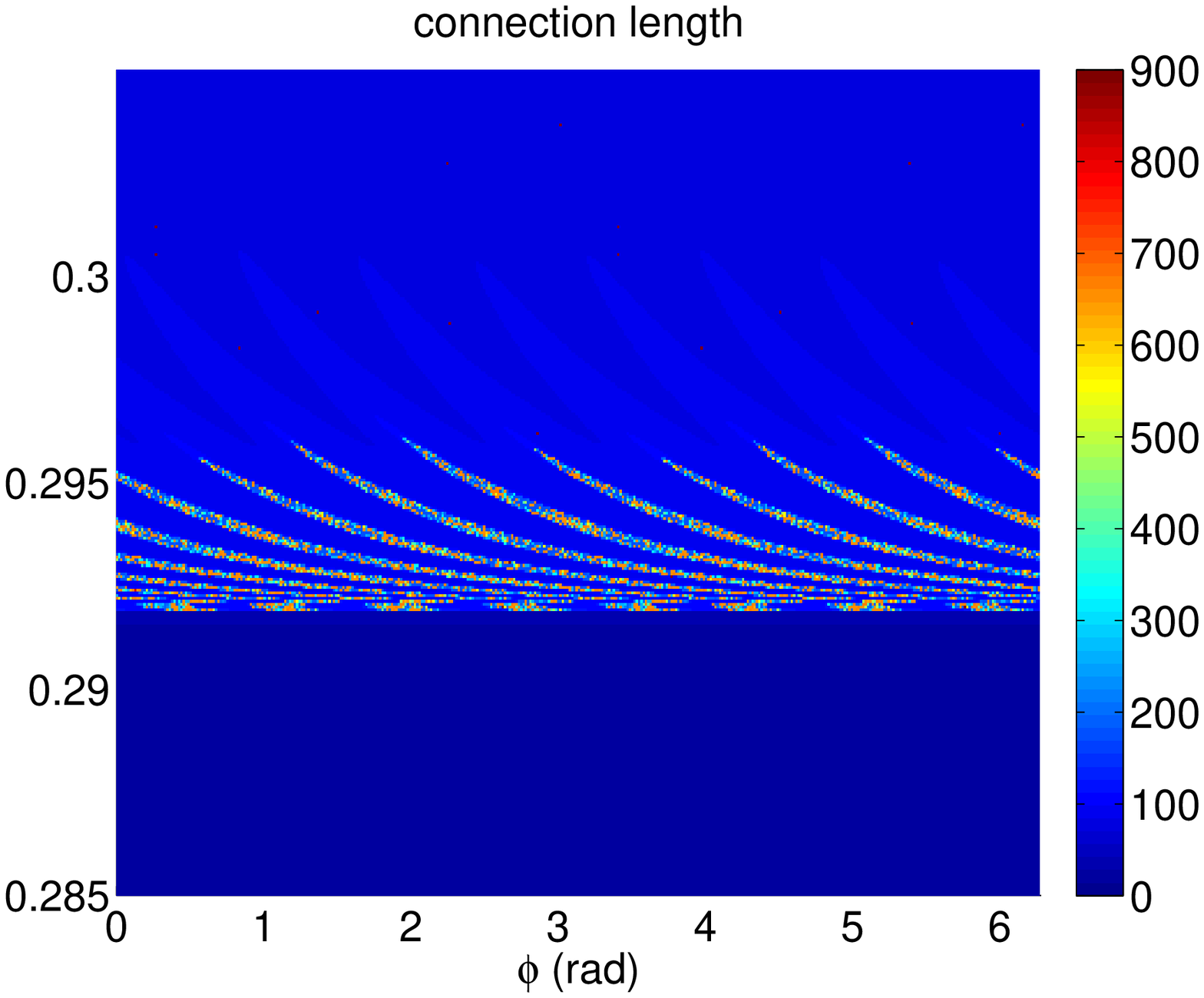}
  \includegraphics[width=50mm]{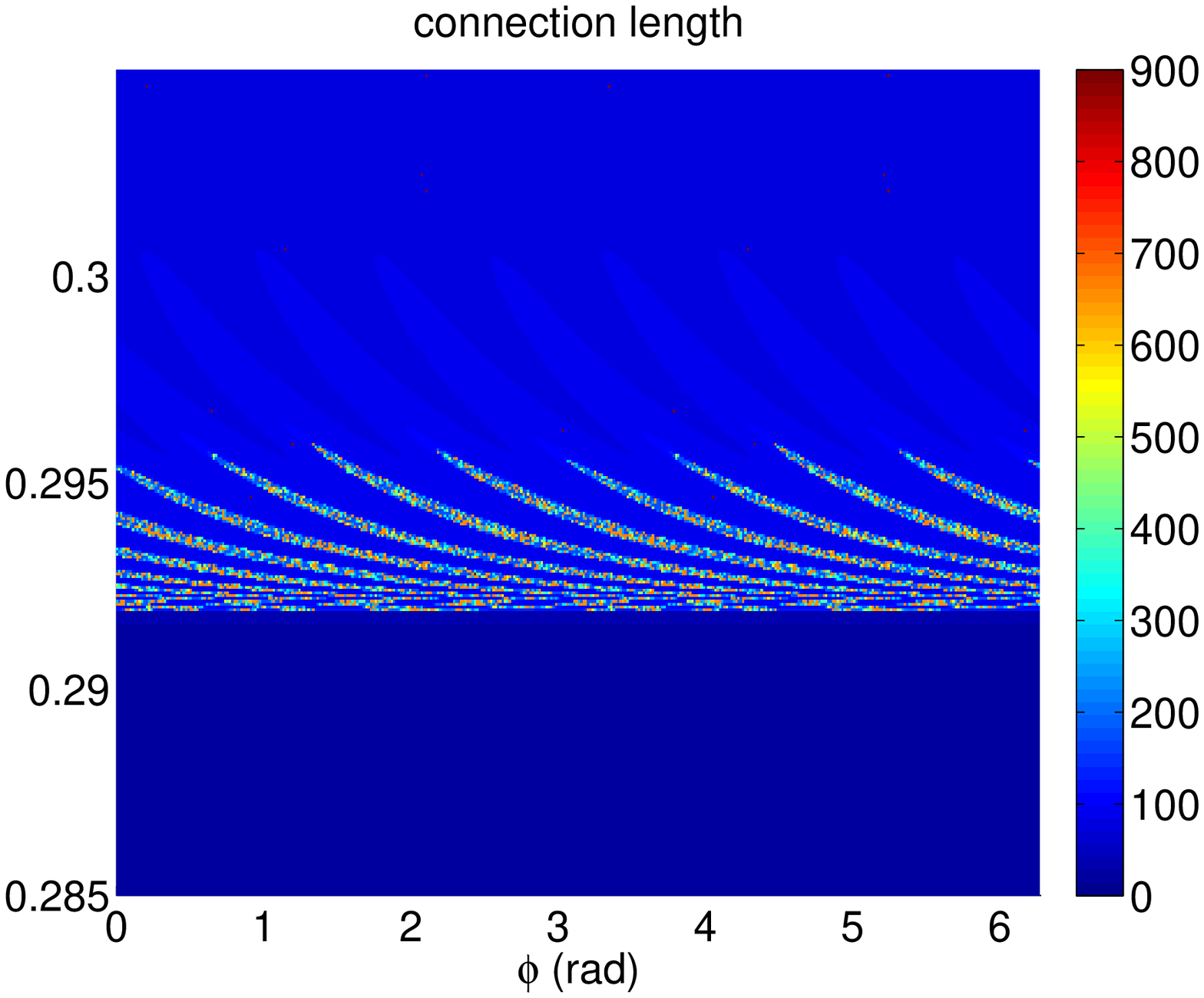}
  \includegraphics[width=50mm]{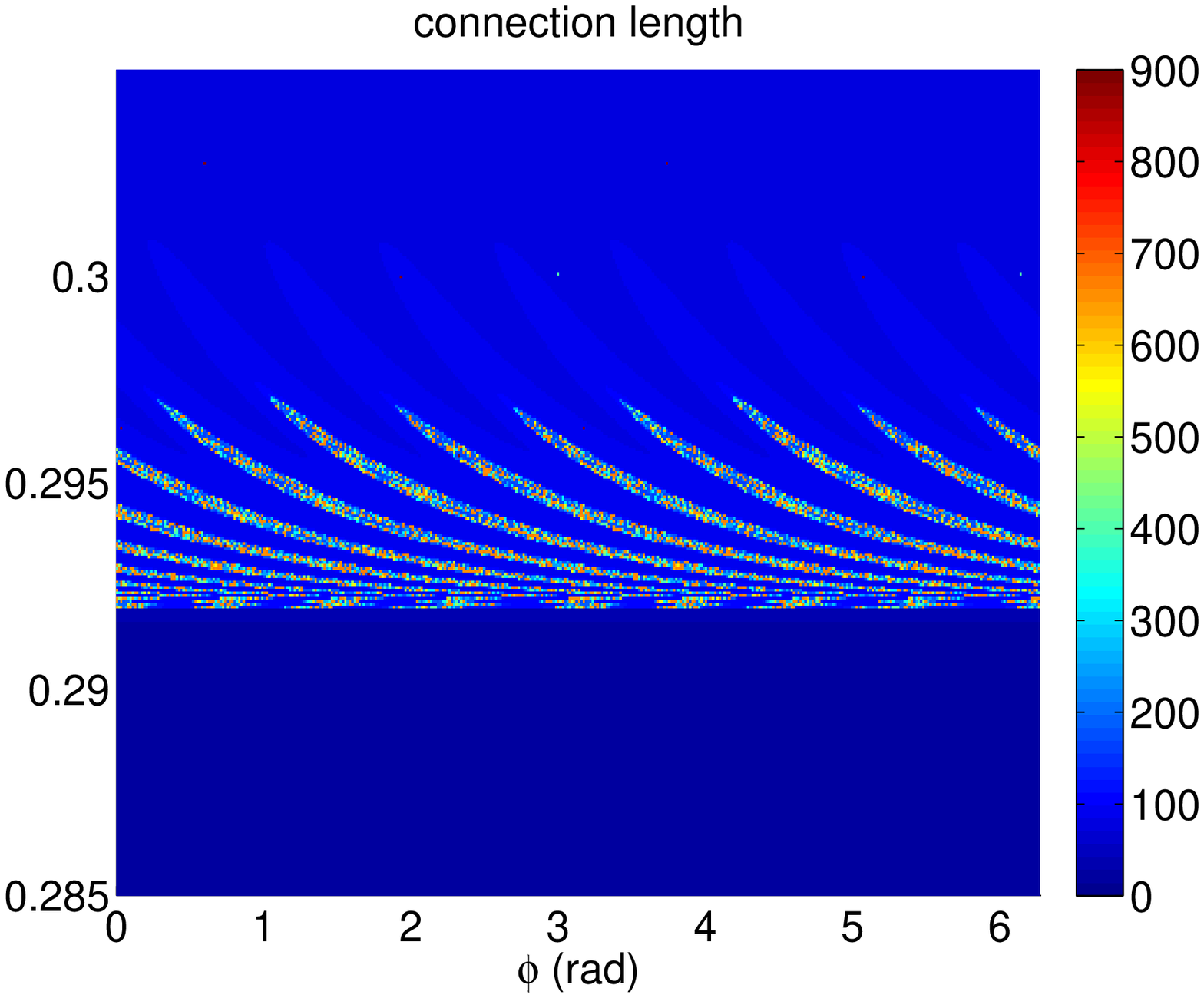}
  }%
  \caption{Connection length on divertor during an ELM crash modelled
    by JOREK in three time instants: $1.1119\cdot10^{-3}$~s, $1.1142\cdot10^{-3}$~s, $1.1165\cdot10^{-3}$~s.}%
 \label{fig:n8ELMonly}%
\end{figure}
figure~\ref{fig:n8ELMonly} shows laminar plots on the divertor
calculated for three time instants during the ELM crash in a JOREK
simulation. The dominant toroidal mode number is $n=8$ here,
corresponding to the 8 visible footprint lobes. We may see that the
whole pattern is rotating.

\begin{figure}[htb]%
  \centering%
  \tiny{\includegraphics[width=50mm]{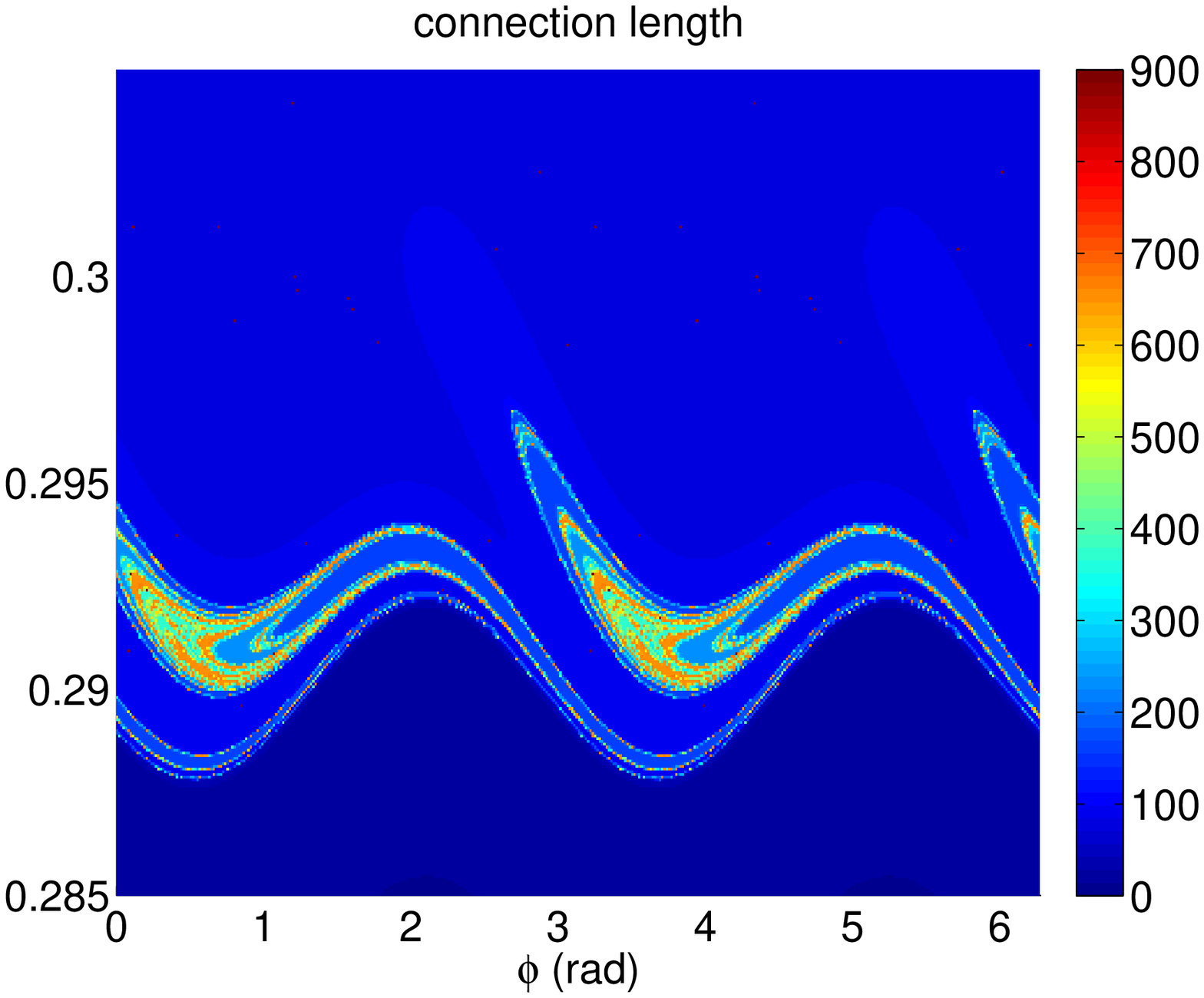}
  \includegraphics[width=50mm]{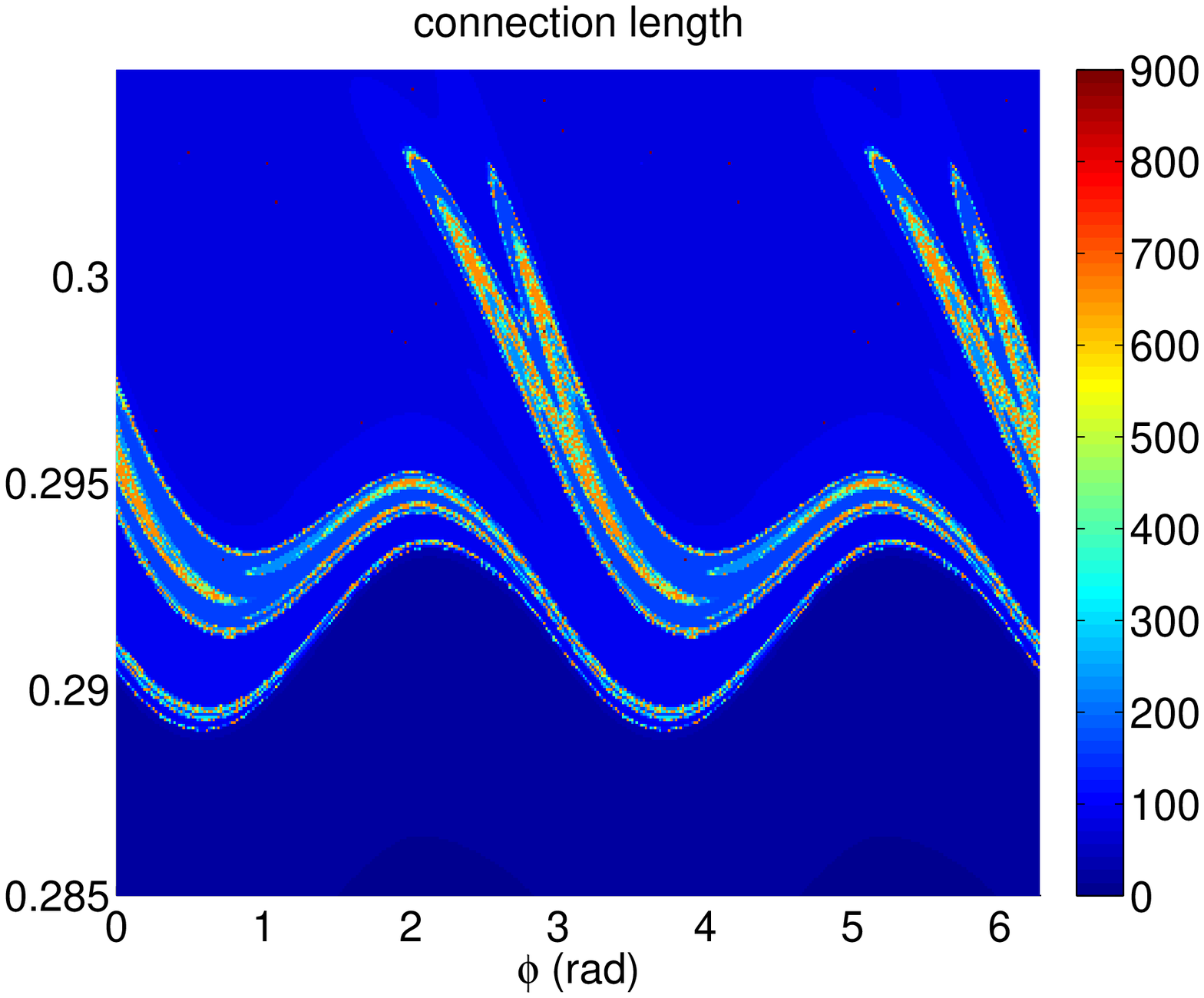}
  \includegraphics[width=50mm]{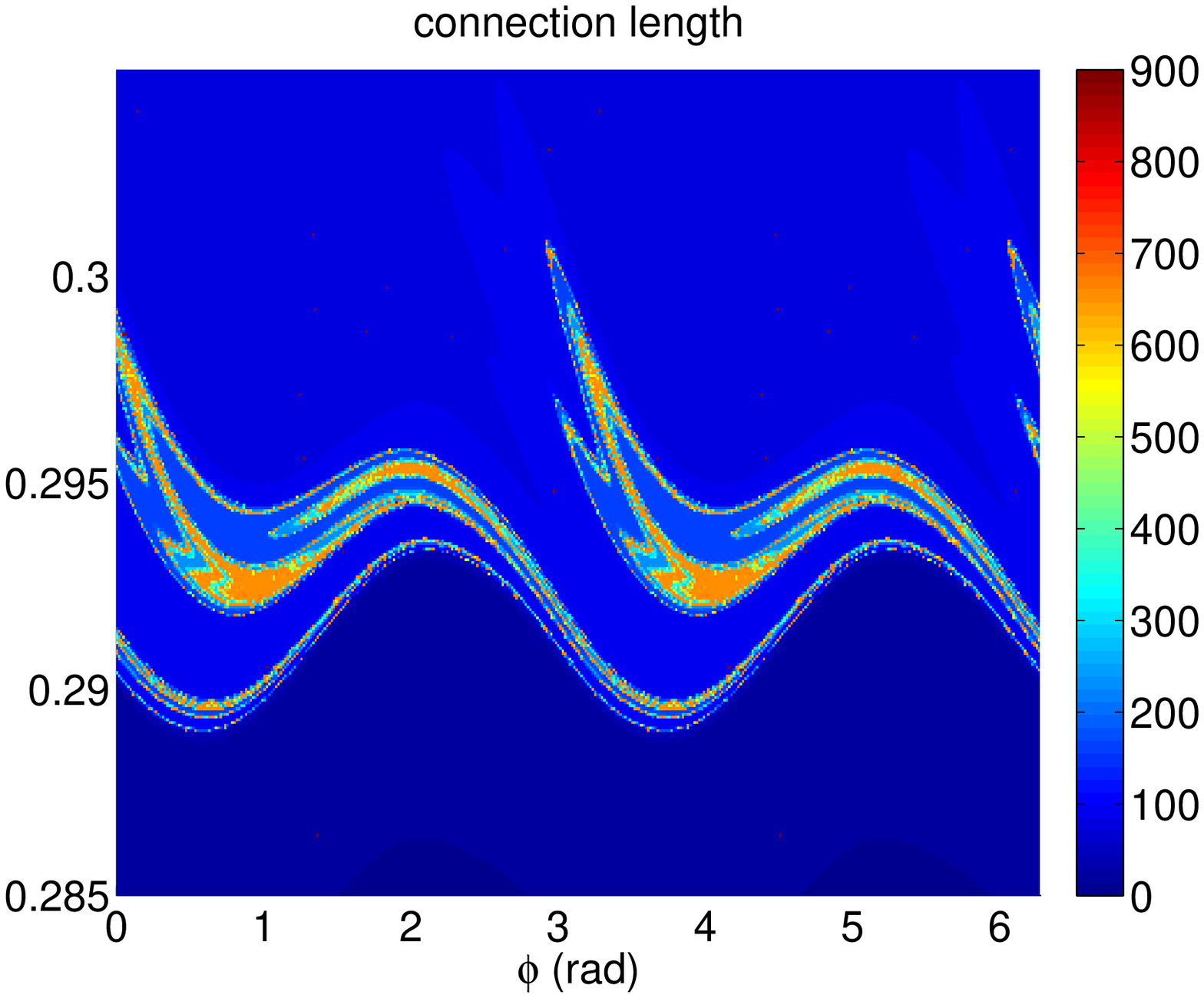}
  }%
  \caption{Connection length on divertor with a $n=2$ RMP (left) and
    together with an ELM crash in two time instants (center:
    $t=2.576\cdot10^{-3}$~s; right: $t=2.601\cdot10^{-3}$~s).}%
 \label{fig:ELMplusRMP}%
\end{figure}

The next simulations were run with a $n=2$ RMP similar to the
one produced by JET EFCCs. Figure~\ref{fig:ELMplusRMP} shows the
resulting laminar plot without ELMs (produced by running the
simulation with $n=0$ and $n=2$ harmonics only) and with an ELM crash
(using the harmonics $n=0 \ldots 8$ with a step of 2). The primary
strike point is distorted into an undulating
pattern due to the perturbation being non-negligible in the X-point
area. The footprint pattern with an ELM shows a similar $n=2$ structure as
without an ELM on a large scale, but its boundary has a more complex
shape with multiple spirals on a smaller scale. The fine scale
structures evolve over time while the large scale $n=2$ footprints
remain mostly the same.

\begin{figure}[htb]%
  \centering%
  \tiny{\includegraphics[width=50mm]{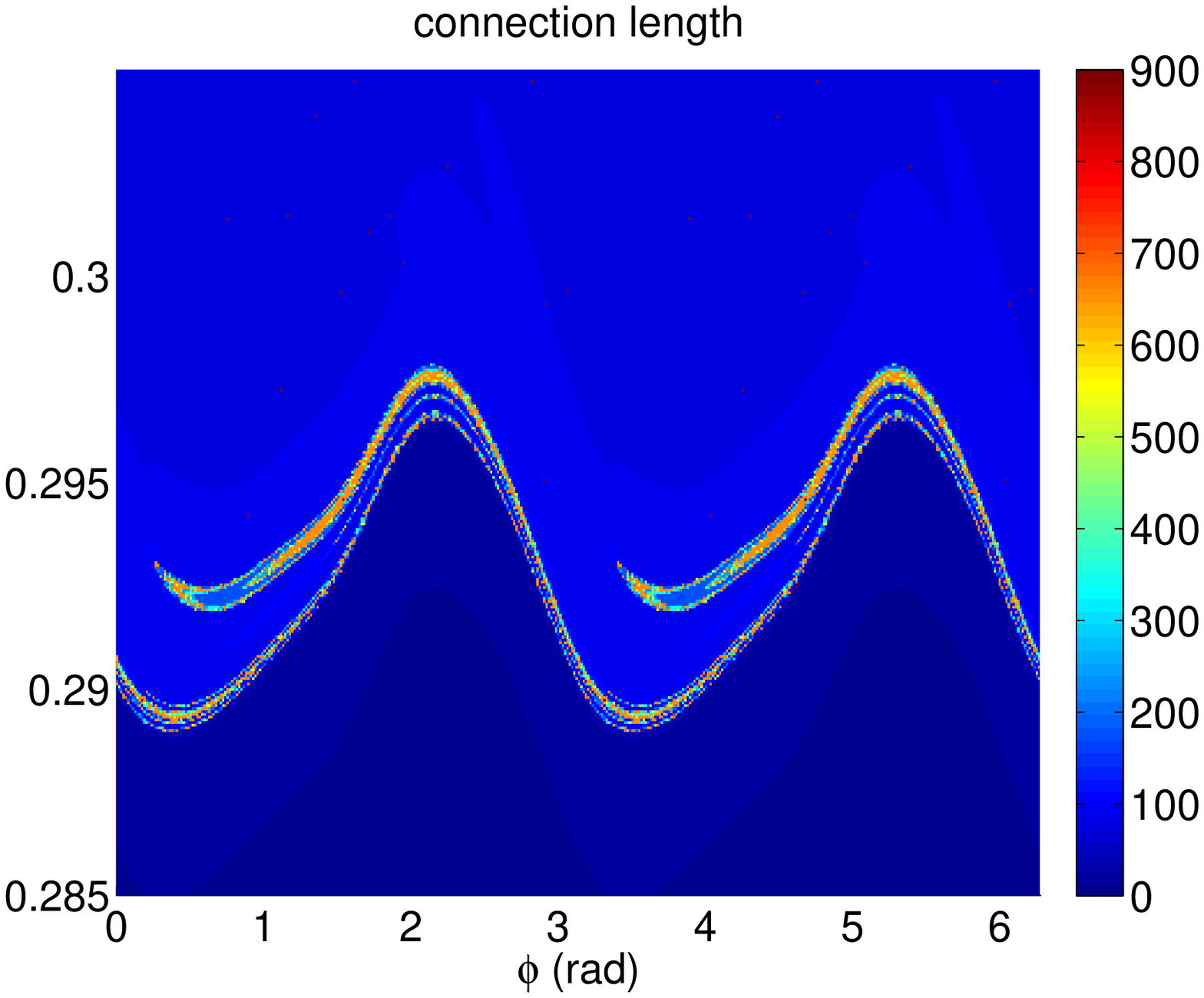}
  \includegraphics[width=50mm]{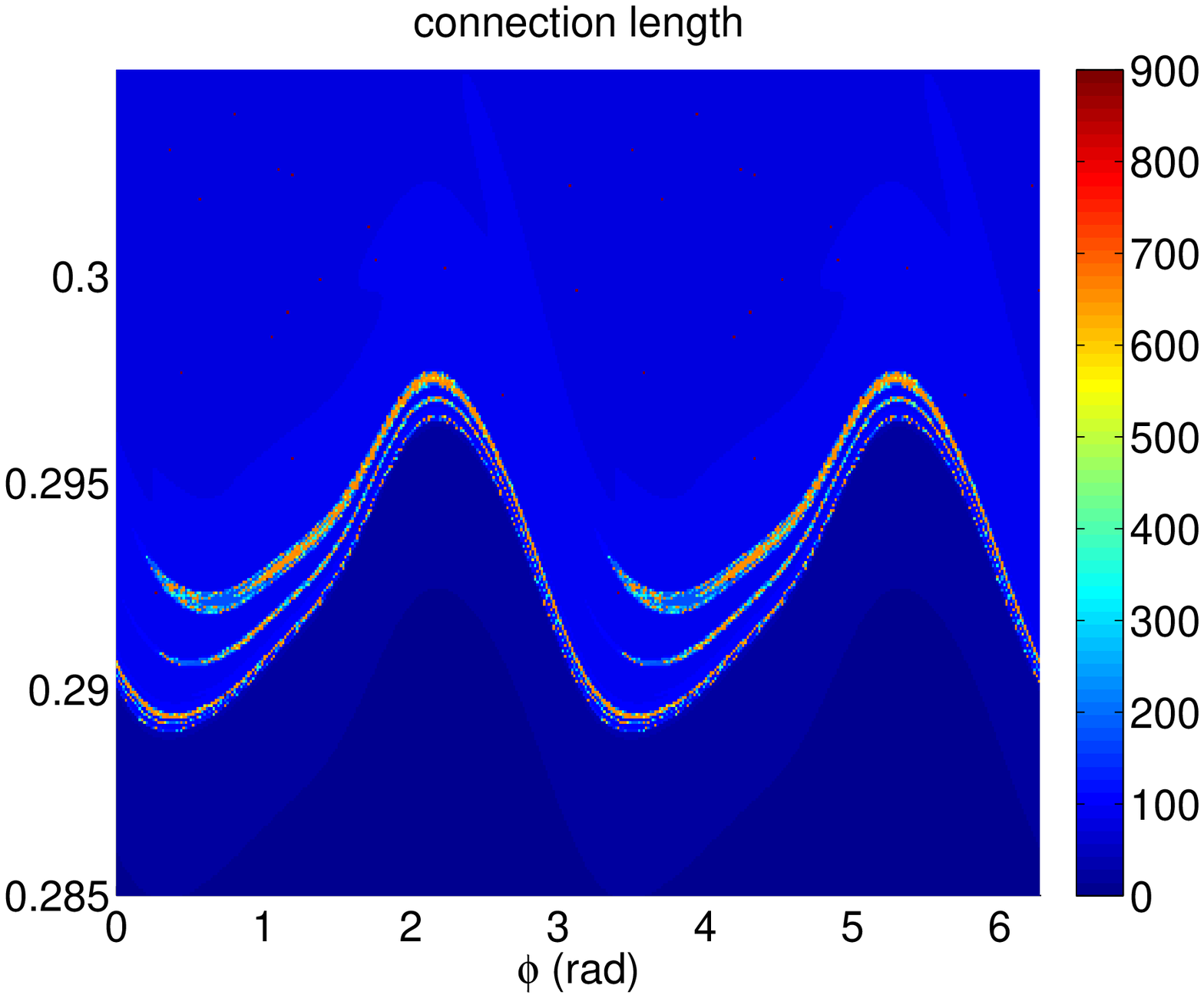}
  \includegraphics[width=50mm]{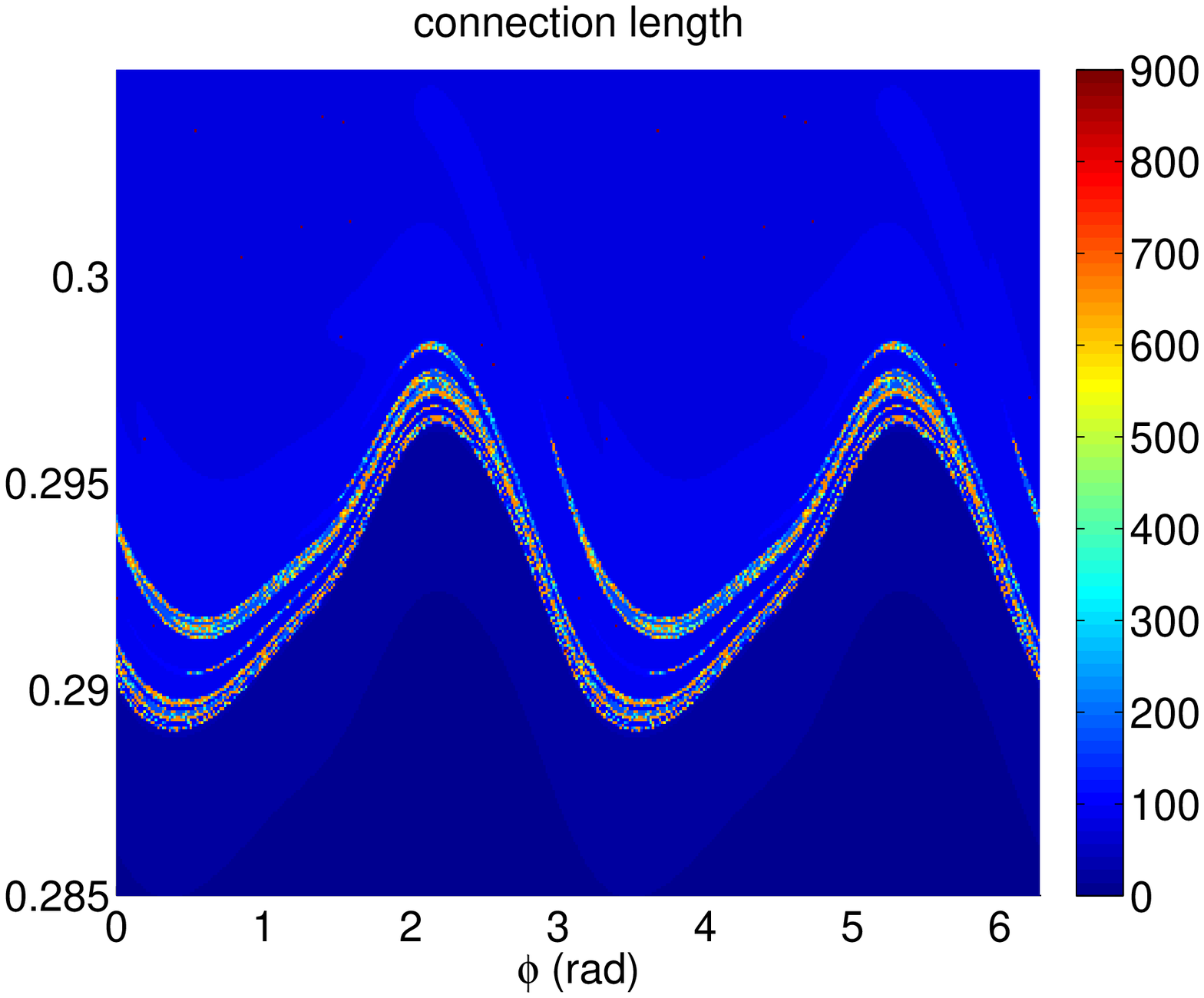}
  }%
  \caption{Connection length on divertor with a $n=2$ RMP and
    an ELM crash with a dominant $n=6$ mode in three time instants:
    $3.5599\cdot10^{-3}$~s, $3.5966\cdot10^{-3}$~s, $3.6332\cdot10^{-3}$~s.}%
 \label{fig:ELMplusRMPn6}%
\end{figure}

It should be noted that while without the RMP the dominant toroidal
mode number of the ELMs is $n=8$, with the RMP the energy ``cascades''
to lower modes $n=4,6$ due to their nonlinear coupling~\cite{:/content/aip/journal/pop/20/8/10.1063/1.4817953} as detailed in~\cite{marinaPRL2014}. We are also interested in
cases where the toroidal mode number is unchanged by the RMP,
corresponding to the experimental observations on MAST. Such a case is
provided by a JOREK simulation with an increased diamagnetic rotation,
where the $n=6$ mode is the most unstable and is not mitigated by RMP
corresponding to a current of 40~kAt in the EFCCs~\cite{marinaiaea2014}. The results for
three time instants are shown in figure~\ref{fig:ELMplusRMPn6}. The
laminar plot has a $n=2$ pattern, split into a structure of finer
footprints which evolve over time while conserving the basic $n=2$
pattern.

\subsection{Footprint rotation}
\begin{figure}[htb]%
  \centering%
  \tiny{\includegraphics[width=50mm]{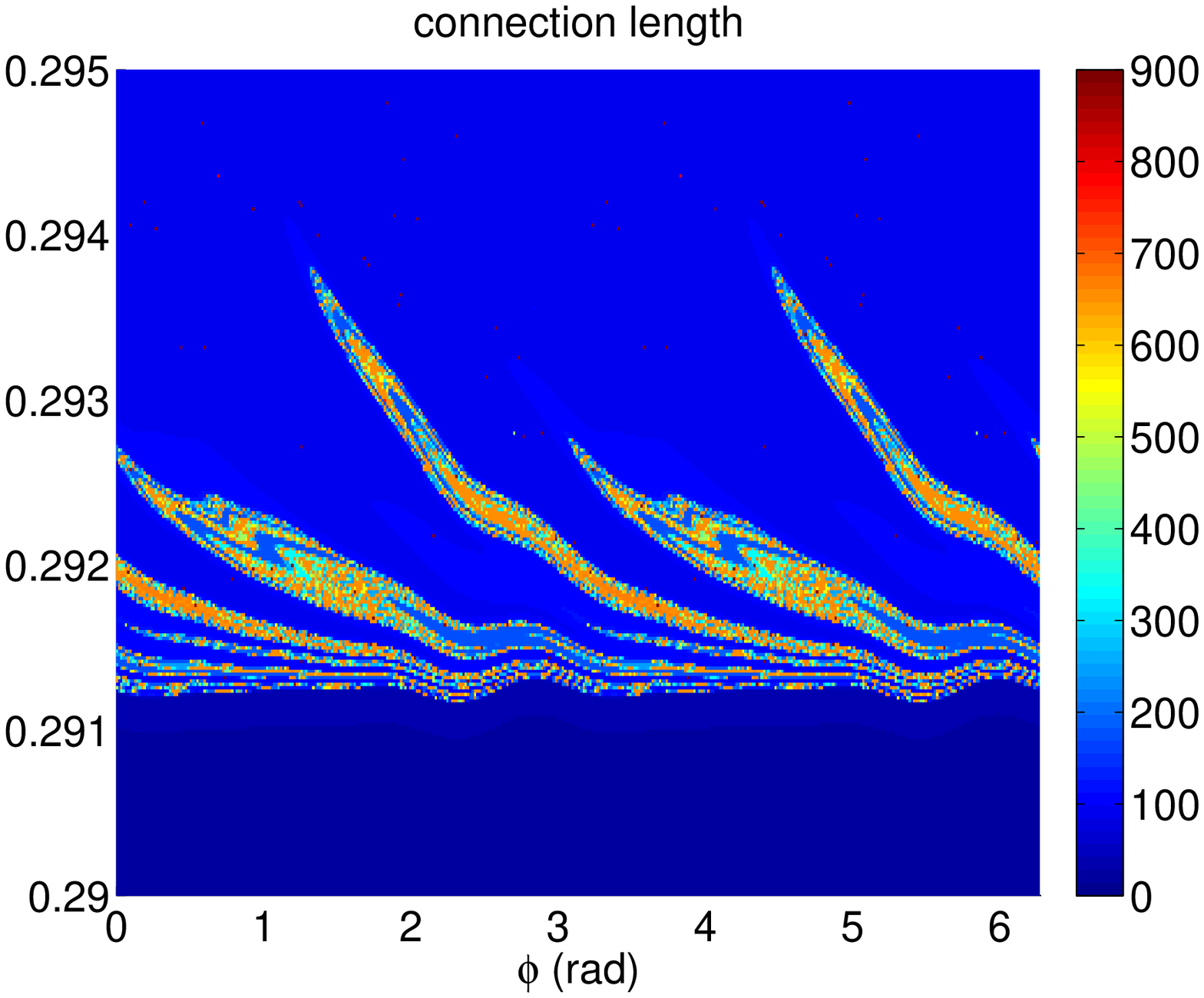}
\includegraphics[width=50mm]{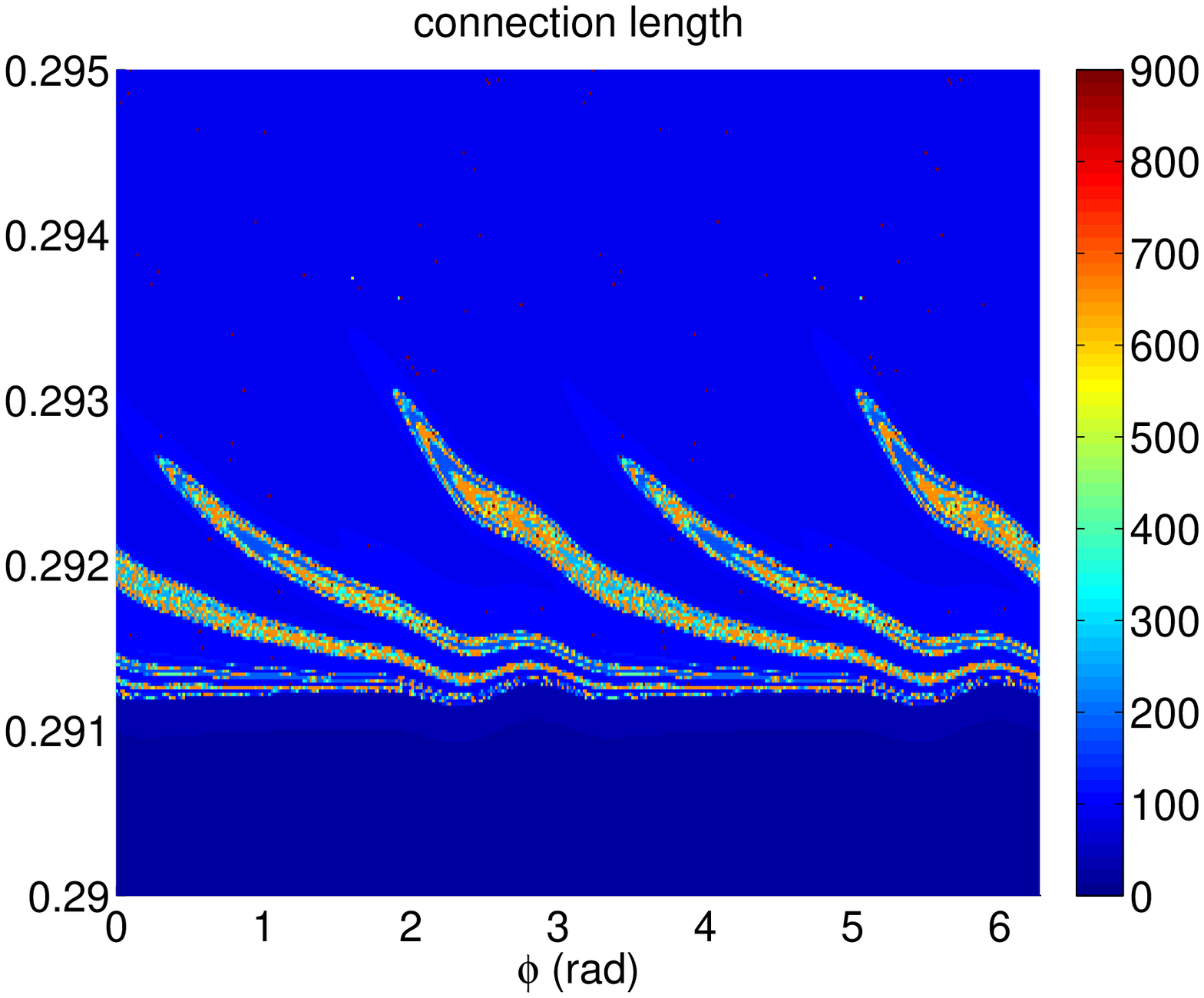}
  \includegraphics[width=50mm]{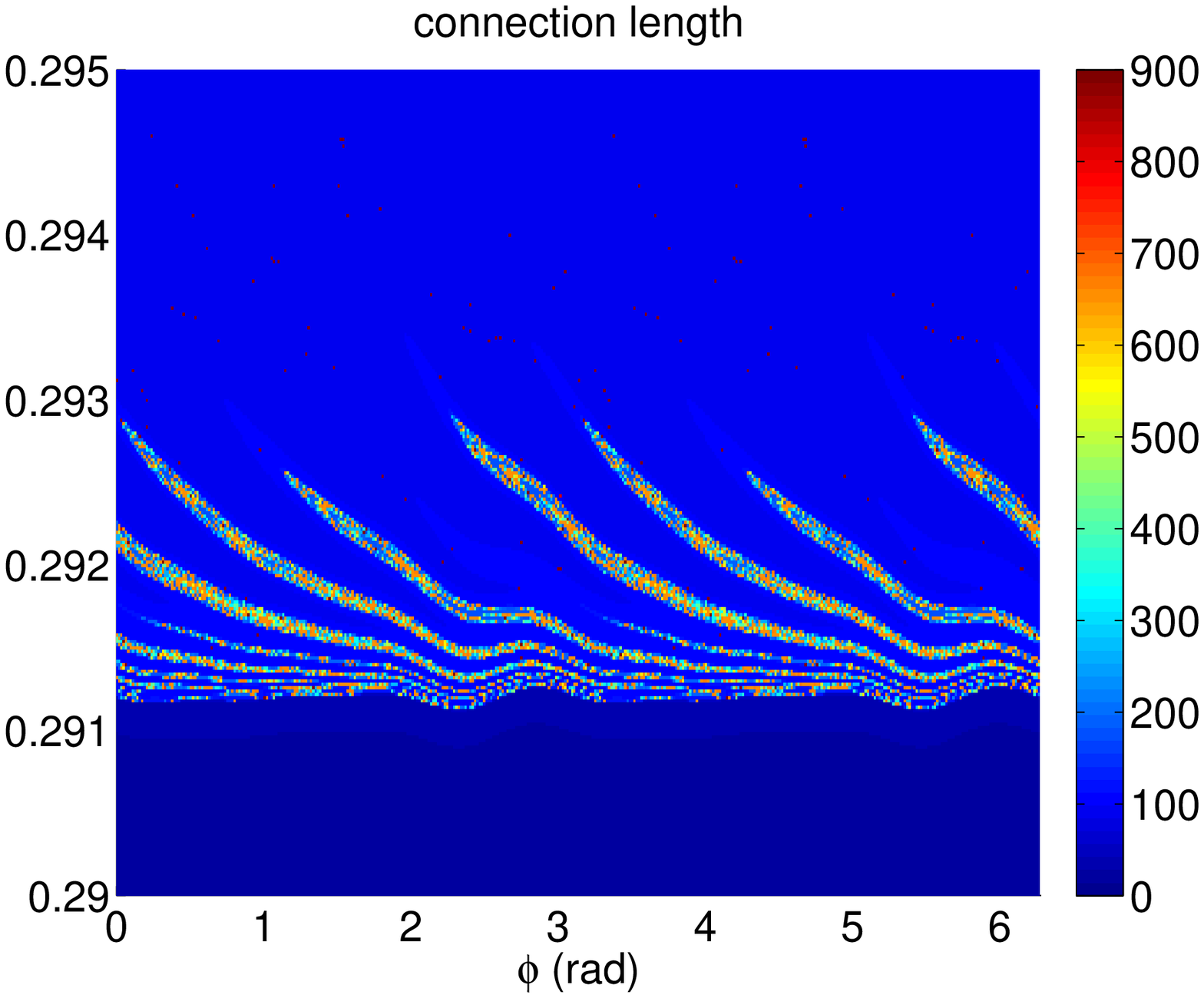}
  }%
  \caption{Connection length on divertor with a mitigated ELM, using
    only $n=0,4,6,8$ components of the magnetic field, in three time instants (left:
    $t=2.576\cdot10^{-3}$~s; center: $t=2.590\cdot10^{-3}$~s, right: $t=2.601\cdot10^{-3}$~s).}%
 \label{fig:ELMplusRMPskipRMP}%
\end{figure}

We may observe that without the magnetic perturbation the footprints
rotate in the direction of the increasing toroidal angle $\phi$. This
is the clockwise direction when viewed from top and is also the
direction of the electron diamagnetic and $E \times B$ drifts, in
which the ballooning instability rotates~\cite{MoralesEPS2014}. With
the magnetic perturbation, the fine details of the boundary fluctuate,
which may be due to the rotation of the higher $n$ modes due to the
ELMs. The superposition with the dominant $n=2$ pattern from the RMP
does not allow to distinguish the rotation clearly, though. For this
reason we performed the field line tracing again in a field
constructed by subtracting the $n=2$ mode from the total field, while
the higher modes are being kept. The $n=2$ mode is still present in
the MHD simulation and thus modifies the other modes via nonlinear
coupling, it is removed only in the field line tracing step. The footprints calculated in this way
are somewhat artificial --- they
do not correspond to power loads on the divertor in any way --- but allow us to
determine the rotation of the higher modes (corresponding to ELMs). The laminar plots
corresponding to figure~\ref{fig:ELMplusRMP} (with one plot added at
an intermediate time instant) are shown in figure~\ref{fig:ELMplusRMPskipRMP}.
The footprint pattern still rotates in the direction of increasing
$\phi$, albeit more slowly than in the unmitigated phase. Due to the
presence of multiple modes, the pattern does not only rotate, but also
progressively changes its shape as the relative amplitudes of the
modes vary. We may also note that the periodic distortion of the
primary strike point due to the RMP has vanished. For comparison,
figure~\ref{fig:ELMplusRMPskipRMP} shows the poloidal profiles of density at the same time
instants (in this case with all the harmonics including $n=2$). The
pattern rotates in the electron diamagnetic direction (counterclockwise).

\begin{figure}[htb]%
  \centering%
  \tiny{\includegraphics[width=50mm]{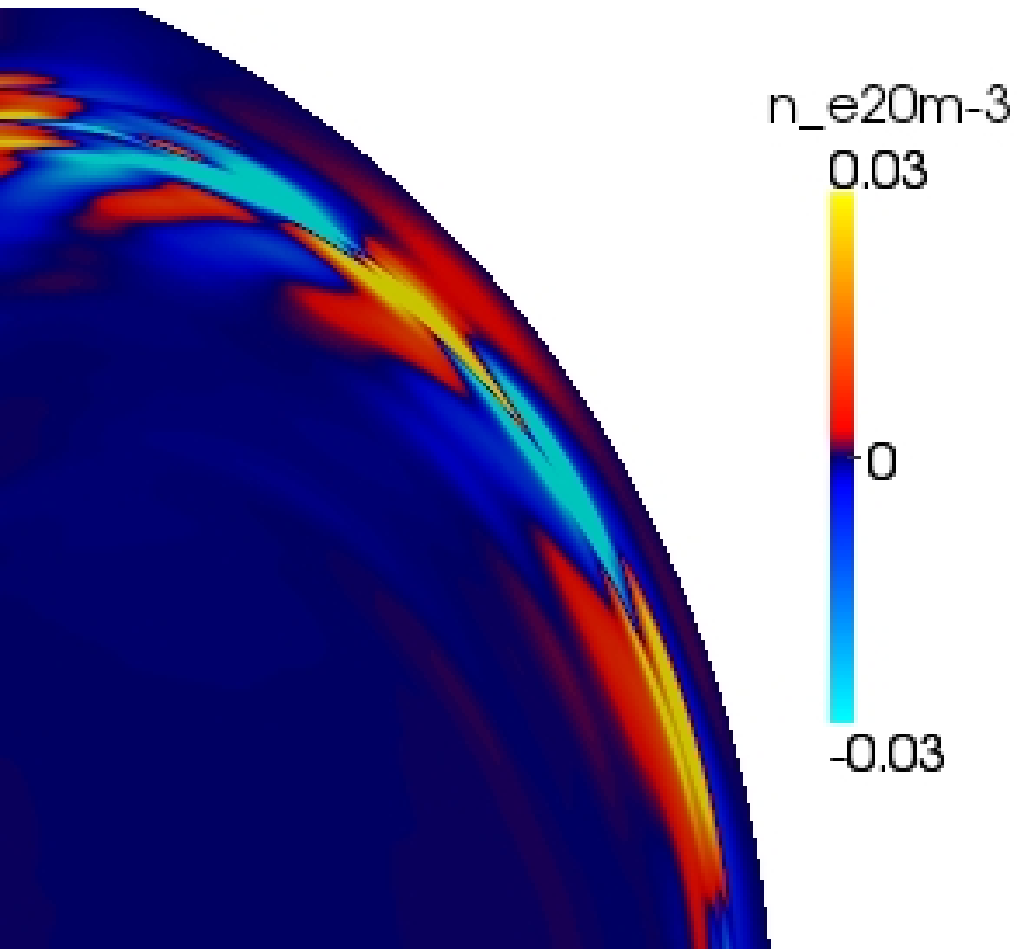}
\includegraphics[width=50mm]{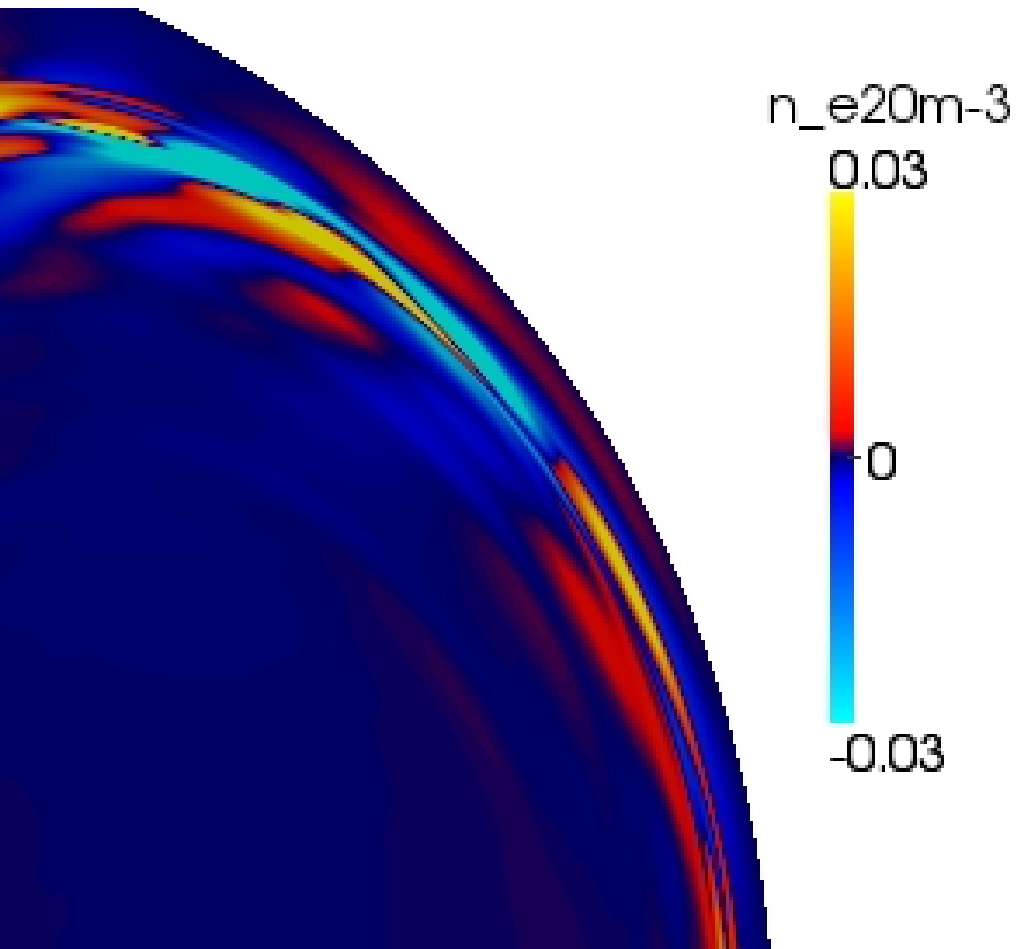}
  \includegraphics[width=50mm]{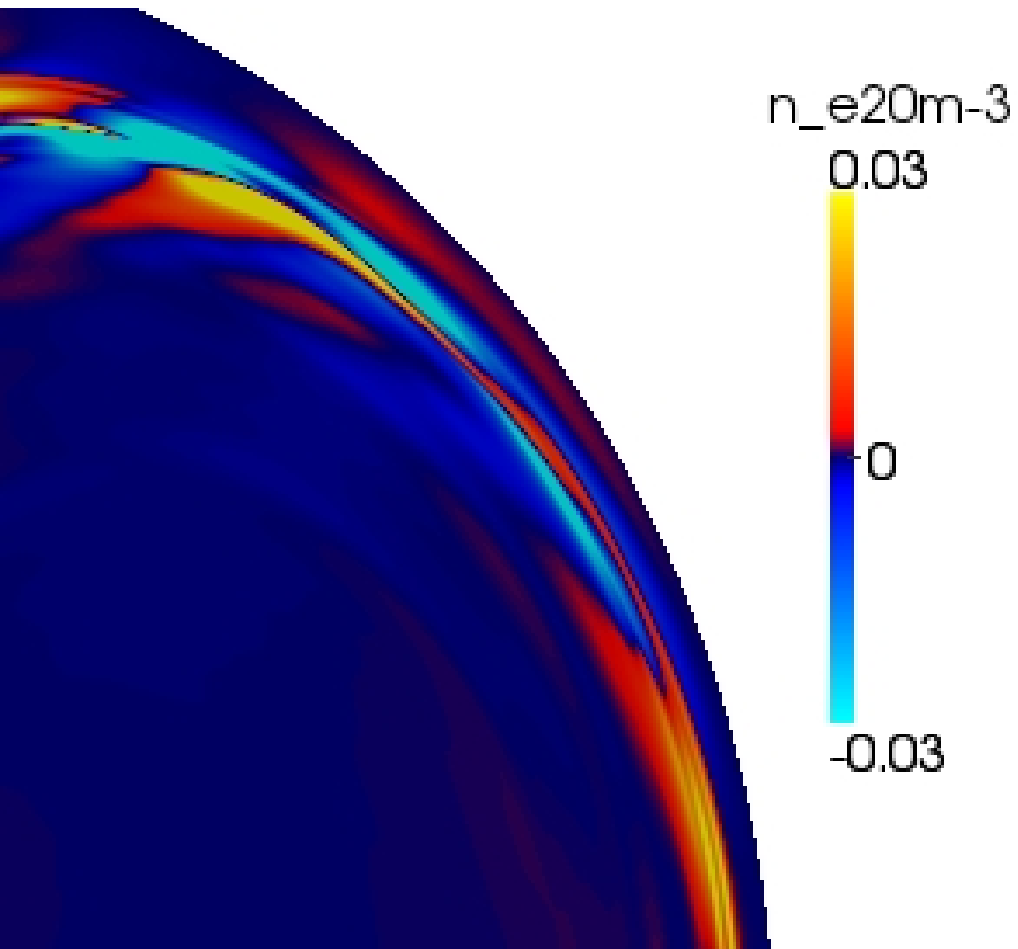}
  }%
  \caption{Density in the poloidal plane during the mitigated ELM crash in three time instants (left:
    $t=2.576\cdot10^{-3}$~s; center: $t=2.590\cdot10^{-3}$~s, right: $t=2.601\cdot10^{-3}$~s).}%
 \label{fig:ELMplusRMPdensitypoloid}%
\end{figure}

\begin{figure}[htb]%
  \centering%
  \tiny{\includegraphics[width=50mm]{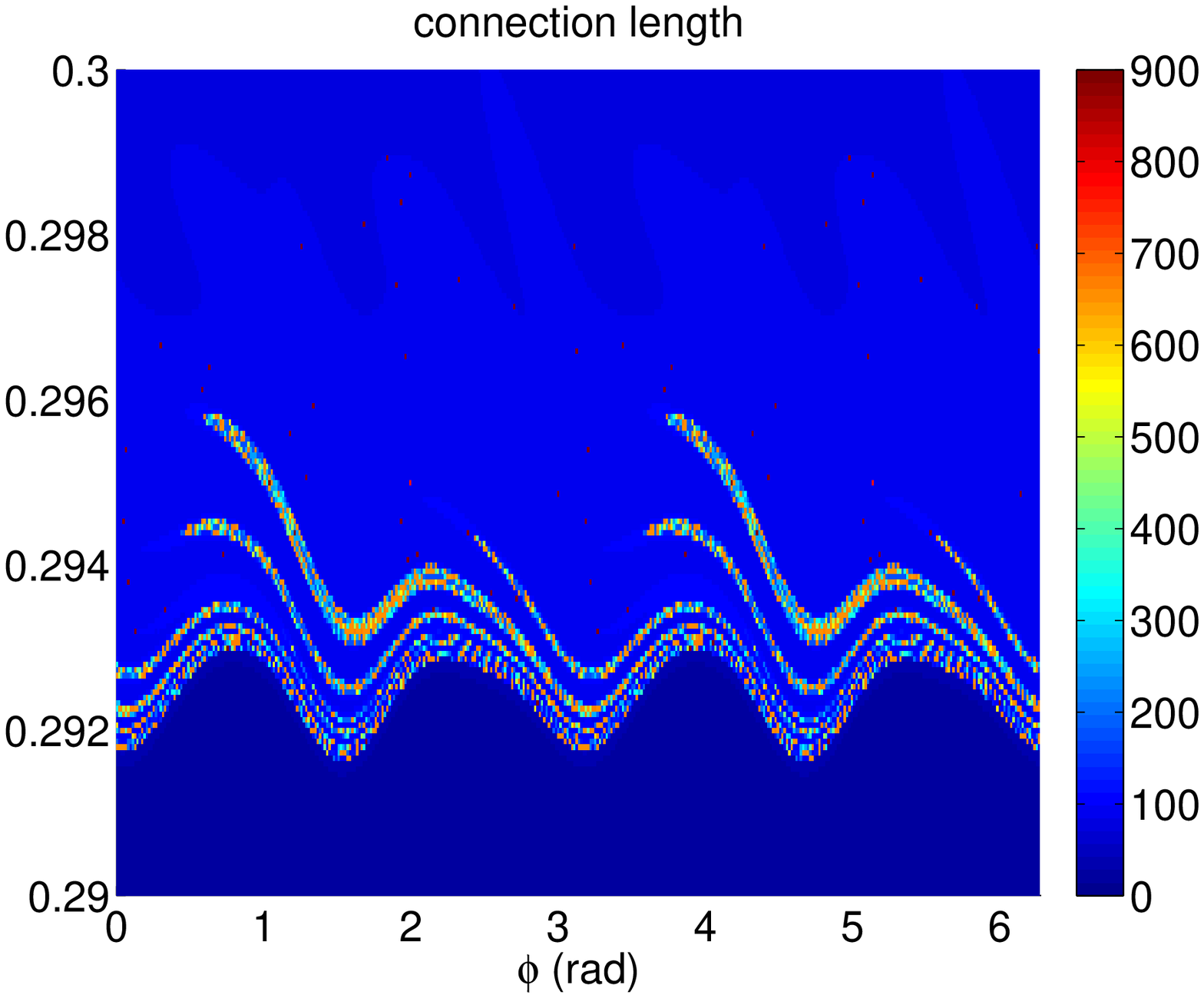}
  \includegraphics[width=50mm]{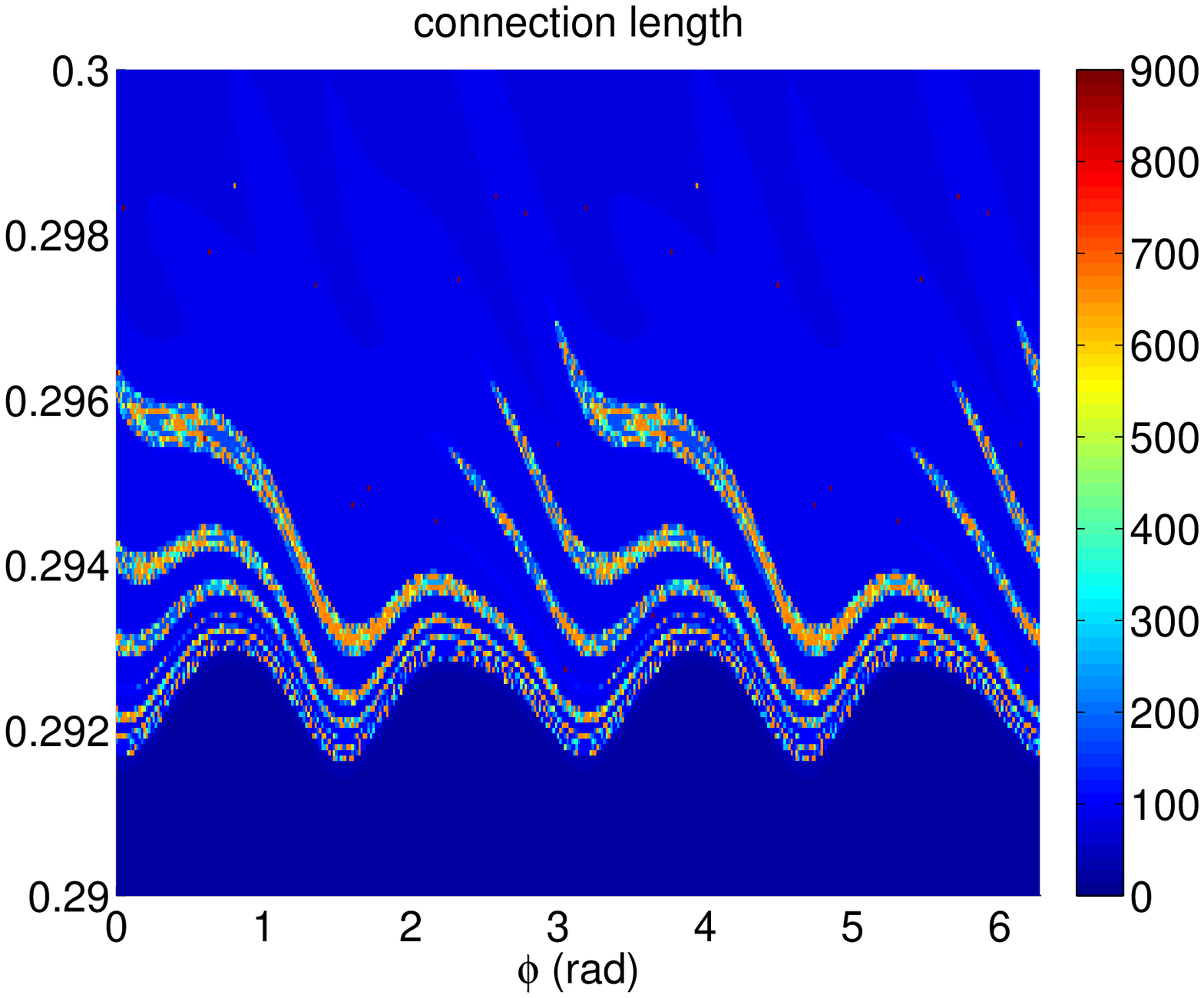}
  \includegraphics[width=50mm]{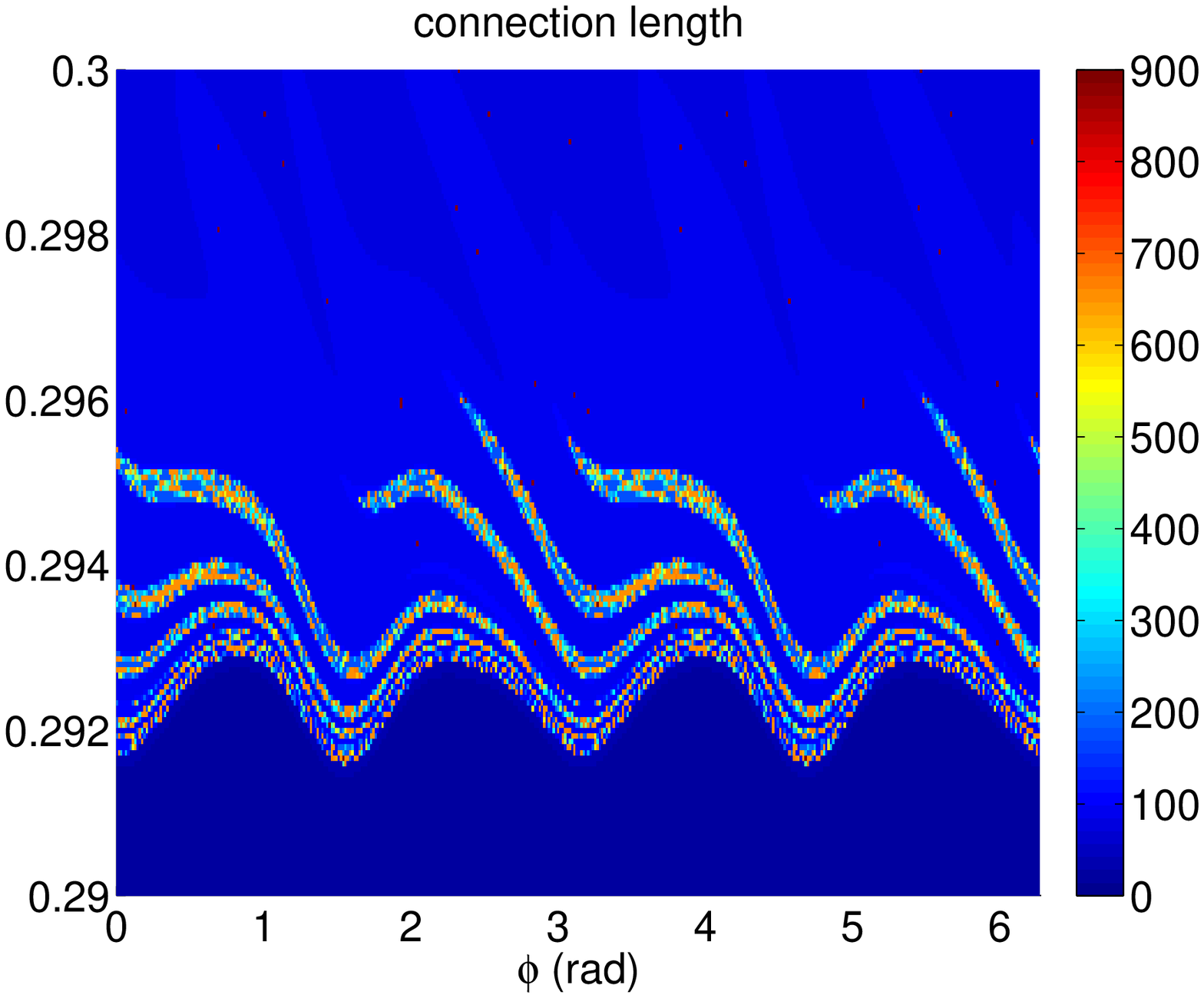}
  }%
  \caption{Connection length on divertor with a $n=2$ RMP and
    an ELM crash with a dominant $n=6$ mode, using
    only $n=0,4,6,8$ components of the magnetic field, in three time instants:
    $3.5599\cdot10^{-3}$~s, $3.5966\cdot10^{-3}$~s, $3.6332\cdot10^{-3}$~s.}%
 \label{fig:ELMplusRMPn6skipRMP}%
\end{figure}

For the case when the RMP does not mitigate
the ELMs (figure~\ref{fig:ELMplusRMPn6}) the corresponding
laminar plot without the $n=2$ mode is shown in figure~\ref{fig:ELMplusRMPn6skipRMP}.
Again, the footprint pattern of the ELM rotates and changes its shape.

\section{Discussion and conclusion}
We have shown that a simple linear superposition of the field of an
ELM and an applied RMP field leads to concentration of open field
lines in the footprints of the RMP field. (Remarkably, our model does
not rely on any nonlinear interaction between the ELM and the RMP,
although such an interaction is known to occur in the MHD
simulations~\cite{marinaPRL2014,:/content/aip/journal/pop/20/8/10.1063/1.4817953}.) Assuming that at least a
part of the ELM loss is caused by parallel transport along field
lines, this would lead to an apparent ``locking'' of the divertor heat
flux pattern to the RMP even though the ELM has a different toroidal
mode number and its rotation may not actually lock. This was indeed observed on MAST, where though the ELM
heat flux follows the RMP footprints only when averaged over the start
of the ELM pulse. This result can be reconciled with our model by
using the fact that a part of the ELM energy is carried away by
filaments and only the rest may be caused by the transport in the
homoclinic tangle assumed here. Furthermore, the filaments arrive
at the target with a delay~\cite{0741-3335-49-8-011}. If the structure of the filament
losses is unaffected by the RMP, they will in a later stage of the ELM
crash blur the pattern caused by the stochastic transport in the
beginning, explaining the observed effect. 
We shall note that only the
structure of the stochastic losses was studied here and further
research is needed to determine if the RMP has any impact on the
filament losses or not. We also studied only the connection length in
the JOREK simulations, as a too high perpendicular diffusion needed
for numerical stability reasons prevented us from extracting
meaningful target power profiles. Solving this problem will allow us
to tackle also the filament question.

While our theory predicts that the footprints will have generally the
shape of the RMP footprints, the ELM field causes an additional distortion of
their edge, depending on the relative amplitude of the ELM compared to
the RMP. If the footprints due to the ELM field are longer than those
due to the RMP, the RMP structure may even stop being distinguishable.
This may explain the NSTX results where the RMP structure was seen
only in smaller ELMs~\cite{0741-3335-56-1-015005}, assuming that
larger ELMs have longer footprints. The distortion of the footprints was  seen especially in the $n=6$ JOREK runs, where
the ELMs are not mitigated. This fine distortion is moving across the
static RMP pattern in the JOREK simulations. The modification of the
static RMP pattern was proposed in~\cite{0029-5515-54-6-064011} as the
explanation of MAST results where the ELM load follows the footprints
only in some ELMs. If the phase of the ELM is such that its Melnikov
integral has a constructive interference with the one of the RMP at
the footprint tip, the footprint will become longer and thus better
observable. On the other hand, a destructive interference at the tip
will reduce it. In~\cite{0029-5515-54-6-064011} it was noted that this
explanation requires that the ELM phase locks (i. e., does not
rotate). This is a different regime than observed in the JOREK
results where the phase always rotates. The rotation was observed to
slow down though in our simulations, so if the change of phase is
negligible during the infra-red camera exposure time (35~$\mu s$  on MAST), the result
may be indistinguishable from complete locking of the phase. In this
way the experimental observations may be reconciled with the simulations.

Our model relies only on the general assumption that an ELM has
a magnetic component which can form a homoclinic tangle, and does not
depend on a particular origin of this field. We have seen that JOREK
simulations predict the existence of such a field during the ELM
crash. Another suggestion was that such a field is generated by
thermoelectric currents flowing through the homoclinic tangle~\cite{PhysRevLett.104.175001}. We
may say that this theory is not needed for a qualitative explanation
of the experimental results, as JOREK simulations do not include the
thermoelectric currents and still show the effect. It may be the case
though that a quantitative analysis will show the need for the
thermoelectric currents in order to explain the data. Such a claim was
made in~\cite{0029-5515-54-6-064012}, where it was found that the vacuum RMP footprints do not
match those observed during the mitigated ELMs on JET. The model used
in~\cite{0029-5515-54-6-064012} did not apparently include the field of the ELM itself, so the
possibility that the vacuum RMP field superposed with the ELM field will
be sufficient to match the data can not be completely ruled out yet.

Further work shall concentrate on determining the changes of the ELM
wetted area due to the effect described here. An increase of
the wetted area due to RMPs was reported~\cite{0029-5515-53-7-073036}
on JET with the ITER-like wall (contrary to the carbon wall results)
and attributed to the observed footprint pattern, while on MAST the
wetted area was found to
decrease~\cite{0741-3335-55-1-015006,0741-3335-55-4-045007,Thornton2013S199}. The
latter effect is probably not due to the footprint pattern, as it
follows the same scaling as the unmitigated ELMs. Other possible
mechanisms behind the footprint pattern in mitigated ELMs should be
investigated. For instance, if an ELM has a toroidal mode number equal to the one of
the RMP, it may directly lock to it, as suggested
in~\cite{0029-5515-49-9-095013}, and produce a large pattern due to
constructive interference. The influence of RMPs on the filament
pattern shall be modeled, as filaments are responsible at least for a
part of the ELM loss.

\ack 
This work was  part-funded by the
Czech Science Foundation under grants GA14-35260S and
GAP205/11/2341. This work was granted access to the HPC resources of
Aix-Marseille Universit\'e financed by the project Equip@Meso
(ANR-10-EQPX-29-01) of the program  \guillemotleft Investissements d'Avenir\guillemotright{} supervised by the Agence Nationale pour la Recherche.
This work has been carried out within the framework of the EUROfusion
Consortium and has received funding from the European Union's Horizon
2020 research and innovation programme under grant agreement number
633053. The views and opinions expressed herein do not necessarily
reflect those of the European Commission.

The views and opinions
expressed herein do not necessarily reflect those of the ITER Organization.

\section*{References}

\end{document}